\documentclass[%
 aip,
 amsmath,amssymb,
preprint,%
]{revtex4-1}

\usepackage{graphicx}% Include figure files
\usepackage{dcolumn}% Align table columns on decimal point
\usepackage{bm}% bold math
\usepackage{multirow}
\usepackage[utf8]{inputenc}
\usepackage[T1]{fontenc}
\usepackage{mathptmx}

\usepackage{braket}
\usepackage{graphicx}% Include figure files
\usepackage{amsmath,amssymb,mathtools}
\usepackage{dcolumn}% Align table columns on decimal point
\usepackage{bm}%bold math
\usepackage{subfigure}
\usepackage{color}
\begin{document}

\preprint{AIP/123-QED}

\title{Accurate Singlet-Triplet Gaps in Biradicals via the Spin Averaged Anti-Hermitian Contracted Schr\"odinger Equation}% Force line breaks with \\

\author{Jan-Niklas Boyn}
 \affiliation{The James Franck Institute and The Department of Chemistry, The University of Chicago, Chicago, Illinois 60637 USA}%Lines break automatically or can be forced with \\
\author{David A. Mazziotti}%
 \email{damazz@uchicago.edu}
\affiliation{The James Franck Institute and The Department of Chemistry, The University of Chicago, Chicago, Illinois 60637 USA}%

\date{Submitted January 21, 2021\textcolor{black}{; Revised March 13, 2021}}% It is always \today, today,

\begin{abstract}
The accurate description of biradical systems, and in particular the resolution of their singlet-triplet gaps, has long posed a major challenge to the development of electronic structure theories. Biradicaloid singlet ground states are often marked by strong correlation and hence, may not be accurately treated by mainstream, \textcolor{black}{single-reference} methods such as density functional theory (DFT) or coupled cluster theory (CC). The anti-Hermitian contracted Schr\"odinger equation \textcolor{black}{(ACSE)}, whose fundamental quantity is the 2-electron reduced density matrix rather than the $N$-electron wave function, has previously been shown to account for both dynamic and strong correlation when seeded with a strongly correlated guess from a complete active space (CAS) calculation. Here we develop a spin-averaged implementation of the ACSE, allowing it to treat higher multiplicity states from CAS input without additional state preparation. We apply the spin-averaged ACSE to calculate the singlet-triplet gaps in a set of small main group biradicaloids, as well as the organic four electron biradicals trimethylenemethane (TMM) and cyclobutadiene, and naphthalene, benchmarking the results against other state of the art methods reported in the literature.
\end{abstract}

\maketitle

\section{Introduction}
The accurate theoretical description of open-shell and biradical systems is quintessential in the understanding of chemical processes and properties in various areas of modern chemistry. Not only are biradicals fundamental to chemical reactions, where the breaking and formation of chemical bonds passes through a transition state or \textcolor{black}{intermediate} via the unpairing of electrons\cite{birad}, \textcolor{black}{which may lead to significant biradical character}\cite{Rxn1, Rxn2, Rxn3, Rxn4}, they also play roles in catalysis, e.g. as carbene species which have recently garnered significant research interest\cite{carbenes,NCarb}, or in solid state and materials application such as molecular magnets, qubits or photochemical devices\cite{Birad1,Birad2,Birad3}. While modern developments in electronic structure theory have made great advances in the treatment of closed shell molecular systems, allowing common black-box methods such as DFT to be used in the calculation of various electronic properties by chemists from all disciplines with great accuracy, the treatment of biradicals and open-shell systems continues to pose major challenges\cite{DFT1, DFT2, challenges, DFTfundamentalissues}.  \\

\textcolor{black}{Singlet} bi-radicals are perhaps the simplest example of a strongly correlated system: two electrons are distributed in degenerate (or near-degenerate) orbitals yielding a multi-determinate wave function which cannot be expressed in the form of a single Slater determinant, \textcolor{black}{or in the reduced density matrix (RDM) picture, yielding 1-RDMs characterized by fractional, (near-) degenerate natural occupation numbers (NON), or diagonal terms in the 1-RDM}. And while many of the most advanced and commonly used electronic structure methods, such as \textcolor{black}{coupled cluster with singles, doubles and perturbative triples (CCSD(T))} or density functional theory (DFT) resolve dynamic correlation with reliable accuracy, they fail to account for the strong correlation arising from multi-reference states\cite{correlation, correlation2, FSDFTerror}. Traditional methods that do account for static correlation, such as complete active space self consistent field (CASSCF) calculations in turn fail to resolve the dynamic correlation\cite{CASSCF}. It is imperative for new theories to correctly resolve both the static and dynamic components of the total correlation energy\cite{challenges} to obtain an accurate picture of the electronic structure of bi-radicals and open shell systems, and to elucidate successfully their experimentally significant properties, such as their spin-state splittings.  \\

In the last two decades many theories have been presented with the aim to calculate accurately the singlet-triplet gaps of biradicaloids and they can generally be divided into two flavors: DFT-based methods and wave-function-based methods. DFT, \textcolor{black}{while an exact theory}, is inherently single-reference, meaning its calculations rely on a single Slater determinant, and its functionals are known to be notoriously inaccurate at resolving singlet-triplet gaps in di-radicals\cite{DFTbirad1}. This is often overcome by the use of an unrestricted formalism, which while leading to improvements in the resolution of spin state gaps, comes at the expense of unphysical spin densities and electronic energies\cite{DFTfundamentalissues, UDFT, UDFT2}. Multiple theories have been developed to address these issues in DFT, including particle-particle random phase approximation (pp-RPA)\cite{YangPPrpa, pprpa1, pprpa2, pprpa3}, fractional-spin DFT\cite{VFS, FS1, FS2}, multi-configurational pair-density functional theory (MC-PDFT)\cite{PDFT-MG, PDFT1, PDFT2} and spin-flip approaches\cite{SFnew, UB3LYP, KrylovSFDFT}. While these paths have provided for significant improvements, several issues such as the dependence on the chosen functional and lack of \textcolor{black}{consistent} absolute energies remain. On the wave function end of the spectrum, the most ubiquitous method of CASSCF with a second order M\o ller-Plesset perturbative correction (CASPT2) may yield \textcolor{black}{non-physical lower-bound results to the FCI energy}, and inhibitive computational scaling in larger systems\cite{CASPT2}. To overcome these issues, Krylov has pioneered spin-flip approaches that may be based on DFT, configuration interaction (CI) or CC calculations\cite{Krylov-SF-CCSDT, Krylov-SFCCSD, KrylovSFDFT, SFCI}, \textcolor{black}{Piris has developed natural-orbital functional theories\cite{NOFT-Birad}}, and recently Reichman and Friesner have presented an auxiliary-field quantum Monte Carlo approach for the prediction of singlet-triplet gaps that may use CASSCF or unrestricted DFT trial wave functions with great success\cite{RF-AFQMC}. Nevertheless, many of the most accurate and up-to-date programs rely on significant data manipulation, e.g. spin coupling or manual state selection, for accurate calculations, or, \textcolor{black}{in the case of QMC, require highly parallelized computing architectures not available to many researchers}, and thus remain far from being a "black-box" method easy to use for non-theoretical researchers.\\

In this paper we use the anti-Hermitian contracted Schr\"odinger equation (ACSE) in a spin averaged\cite{PirisSpinAvg, Gidofalvi2005, Perez-Romero1997, Mihailovic1975} implementation to resolve accurately the singlet-triplet gaps in multiple biradical systems. The ACSE allows the calculation of the electronic energy as a linear functional of the two-electron reduced density matrix (2-RDM), resulting in favorable polynomial computational scaling compared to wave function based methods\cite{ACSE1, ACSE2}. While the ACSE when seeded with a non-correlated initial guess 2-RDM resolves dynamic correlation, use of an initial 2-RDM obtained from a strongly correlated CASSCF calculation, allows the calculation of both static and dynamic correlation\cite{ACSEcor}. The ACSE has been demonstrated to calculate accurately the various ground and excited states, total correlation energies, and dissociation curves or various molecular systems\cite{ACSE1,ACSE2, ACSEcor, ACSEexp1, ACSEexp2}. In this paper we implement a spin-averaged implementation of the ACSE, seeded from a CASSCF calculation, to resolve the total correlation energy in the $S = 0$ and $S = 1$ states of a common benchmark set of small main group biradicaloids, and several small organic molecules, namely cyclobutadiene, trimethylenemethane (TMM), naphthalene, and o-, m-, p-benzyne. \\

\section{Theory}
The $N$-electron Schro\"odinger equation may be projected onto the space of all two electron transitions by integrating over the coordinates of all but two electrons, yielding the contracted Schr\"odinger equation (CSE)\cite{CSE}:
\begin{equation}
    \bra{\Psi}\hat{\Gamma}^{i,j}_{k,l} \hat{H}\ket{\Psi} = E\; {}^2D^{i,j}_{k,l} \,,
\end{equation}
where $\hat{\Gamma}^{i,j}_{k,l}$ is a two-body operator whose expectation value yields the elements of the 2-RDM
\begin{equation}
    \hat{\Gamma}^{i,j}_{k,l} =  \hat{a}^{\dagger}_i \hat{a}^{\dagger}_j \hat{a}_l  \hat{a}_k \,,
\end{equation}
and $\hat{H}$ is the Hamiltonian operator
\begin{equation}
    \hat{H} = \sum_{ij} {}^1K^i_j \hat{a}^{\dagger}_i \hat{a}_j + \sum_{ijkl} {}^2V^{ij}_{kl} \hat{a}^{\dagger}_i \hat{a}^{\dagger}_j \hat{a}_l \hat{a}_k \,,
\end{equation}
where ${}^1K$ contains the kinetic and nuclear attraction integrals and ${}^2V$ contains the electron-electron repulsion integrals.
The CSE can be separated into its Hermitian and anti-Hermitian parts:
\begin{equation}
    \bra{\Psi}\{\hat{\Gamma}^{i,j}_{k,l},(\hat{H}-E)\}\ket{\Psi} + \bra{\Psi}[\hat{\Gamma}^{i,j}_{k,l},(\hat{H}-E)]\ket{\Psi} = 0 \,,
\end{equation}
where square brackets denote a commutator and curly brackets the anti-commutator. Selection of only the anti-Hermitian part of the CSE yields the ACSE:
\begin{equation}
    \bra{\Psi}[\hat{\Gamma}^{i,j}_{k,l},\hat{H}]\ket{\Psi} = 0 \,.
\end{equation}
While the full CSE and its Hermitian part contain terms of the 2-, 3- and 4-RDM's, the ACSE only depends on the 1-, 2-, and 3-RDM's; explicit expressions of the CSE and ACSE in terms of the RDM's are outlined in references \cite{ACSE1, ACSE2}. Furthermore, this dependence on the 3-RDM can be resolved by using an approximate reconstruction in terms of the 2-RDM. Several methods of reconstruction have been published and in this work we use the cumulant reconstruction\cite{cumulant, schwinger, CSE35}.
\begin{equation}
    {}^3D^{ij}_{kl} \approx {}^1D^{i}_{q} \wedge {}^1D^{j}_{s} \wedge {}^1D^{k}_{t} + 3 {}^2\Delta^{ij}_{qs} \wedge {}^1D^{k}_{t} \,,
\end{equation}
where
\begin{equation}
    {}^2\Delta^{ij}_{qs} = {}^2D^{ij}_{qs} - {}^1D^{i}_{q} \wedge {}^1D^{j}_{s} \,,
\end{equation}
and $\wedge$ denotes the antisymmetric Grassmann wedge product, which is defined as:
\begin{equation}
    {}^1D^{i}_{k} \wedge {}^1D^{j}_{l} = \frac{1}{2}({}^1D^{i}_{k}{}^1D^{j}_{l} - {}^1D^{i}_{l} {}^1D^{j}_{k}) \,.
\end{equation}
As the 3-RDM terms appear only in the perturbative ${}^2V$ part of the Hamiltonian of the ACSE, this approximate reconstruction of ${}^3D$ neglects the cumulant 3-RDM part of the expansion, approximating ${}^3\Delta^{ijk}_{qst}$ to be zero. \\

The ACSE has successfully been used to study small open shell singlet and high-spin triplet systems, however, the treatment of triplet and higher spin states employs a rather complicated approach requiring the spin coupling of open-shell molecules to form singlet states\cite{ACSEspin}. In particular, this approach involved the spin coupling of hydrogen atom(s) to the open shell wave function to create a singlet state and solving the ACSE in terms of its composite 2-RDM to obtain the composite energy, and finally subtracting the energy of the hydrogen atom to obtain the system energy. It requires careful construction of singlet states through coupling of two or more CASSF reference wave functions of different spin states coupled into a singlet wave function by Clebsch-Gordon coefficients and additional post-processing work to obtain the open-shell 2-RDM and energy from the composite system. In this work we present an approach using spin-averaged RDMs~\cite{PirisSpinAvg, Gidofalvi2005, Perez-Romero1997, Mihailovic1975}, \textcolor{black}{which in general are defined as:}
\begin{equation}
    {}^2 D^{ij}_{kl} = \frac{1}{2S+1} \sum_{M=-S}^{S}\bra{\Psi^{M,S}} \hat{\Gamma}^{i,j}_{k,l} \ket{\Psi^{M,S}} \,,
\end{equation}
or equivalently,
\begin{equation}
     {}^2 D^{i,j}_{k,l} = \frac{1}{2S+1} \sum_{M=-S}^{S} {}^2 D_{i,j;k,l}^{M,S} \,,
\end{equation}
\textcolor{black}{where $S$ and $M$ are the total spin and magnetic or secondary spin quantum numbers, respectively. The use of spin-averaged RDMs results in a streamlined computational implementation that requires no further manipulation of RDMs and enables an easy-to-use program.} \\

Various iterations of the ACSE algorithm have been applied to a range of chemical systems, successfully predicting ground- and excited-state energies, dissociation curves and reaction barriers\cite{ACSE1, ACSE2, ACSEexp1, ACSEexp2, ACSEcor}. The nature of the result obtained from an ACSE calculation depends on the 2-RDM used to initialize the calculation. While use of an HF solution as the initial guess results in resolution of the dynamic correlation, the calculation may also be seeded with a 2-RDM obtained from a CASSCF calculation, allowing the determination of both dynamic correlation and strong correlation effects\cite{ACSEcor}. We hence start with a CASSCF calculation in the desired spin state, carried out with the Maple Quantum Chemistry Package\cite{Maple, QCP}, obtaining strongly correlated 1- and 2-RDMs which are then used as an initial guess in the ACSE. \textcolor{black}{The 3-RDM is constructed from the guess 2-RDM by the cumulant reconstruction outlined above, and not obtained from the CASSCF guess.} \\

While the ACSE requires the 1- and 2-RDM elements in the canonical spin-orbital basis for the $\alpha \alpha$ and $\alpha \beta$ blocks, we obtain initial active space RDMs ${}^1\Tilde{D}$ and ${}^2\Tilde{D}$ in the spatial orbital basis from a CASSCF calculation (the $\alpha$ and $\beta$ denote the +1/2 and -1/2 spin states, respectively). \textcolor{black}{Here the CASSCF calculations are computing the high-spin states, i.e. the $M = +1$ state for the triplet, and the spin trace projects the high-spin 2-RDM onto the spin-free 2-RDM.  The spin-free 2-RDM is then mapped to the spin-averaged 2-RDM by the reconstruction of the spin-dependent $\alpha \alpha$ and $\alpha \beta$ blocks according to}:
\begin{equation}
    {}^1D^{i}_{j} = {}^1D^{\alpha i}_{\alpha j} = {}^1D^{\alpha i}_{\beta j} = \frac{1}{2} {}^1\Tilde{D}^i_j \,,
\end{equation}
\begin{equation}
    {}^2D^{\alpha i, \alpha j}_{\alpha k, \alpha l} = {}^2D^{\beta i, \beta j}_{\beta k, \beta l} =\frac{1}{6} ({}^2\Tilde{D}^{ij}_{kl} - {}^2\Tilde{D}^{ji}_{kl}) \,,
\end{equation}
\begin{equation}
    {}^2D^{\alpha i, \beta j}_{\alpha k, \beta l} = {}^2D^{\beta i, \alpha j}_{\beta k, \alpha l} = \frac{1}{6}(2 \, {}^2\Tilde{D}^{ij}_{kl} + {}^2\Tilde{D}^{ji}_{kl}) \,.
\end{equation}
This is followed by reconstruction of the core and virtual elements of ${}^1D$ and ${}^2D$:
\begin{equation}
    {}^1D_{core} = I , {}^1D_{virtual} = 0 \,,
\end{equation}
\begin{equation}
    {}^2D^{\alpha a, \alpha b}_{\alpha c, \alpha d} = {}^2D^{\beta a, \beta b}_{\beta c, \beta d} = {}^1D^{a}_{c} {}^1D^{b}_{d} - {}^1D^{b}_{c} {}^1D^{a}_{d} \,,
\end{equation}
\begin{equation}
    {}^2D^{\alpha a, \beta b}_{\alpha c, \beta d} = {}^2D^{\beta a, \alpha b}_{\beta c, \alpha d} = {}^1D^{a}_{c} {}^1D^{b}_{d} \,,
\end{equation}
\begin{equation}
    {}^2D^{\alpha p, \alpha q}_{\alpha s, \alpha t} = {}^2D^{\beta p, \beta q}_{\beta s, \beta t} = {}^2D^{\alpha p, \beta q}_{\alpha s, \beta t} = {}^2D^{\beta p, \alpha q}_{\beta s, \alpha t} = 0 \,,
\end{equation}
where $a,b,c,d$ denote core orbitals and $p,q,s,t$ denote virtual orbitals.

Using the spin-averaged 1- and 2-RDMs, we solve the ACSE via a system of differential equations\cite{ACSEextrap}:
\begin{equation}
\begin{aligned}
    E(\lambda + \epsilon) &= \bra{\Psi(\lambda)}e^{-\epsilon S(\lambda)} \hat{H} e^{\epsilon S(\lambda)}\ket{\Psi(\lambda)} \\
    &= E(\lambda) + \epsilon \bra{\Psi(\lambda)}[\hat{H}, \hat{S}(\lambda)]\ket{\Psi(\lambda)} + O(\epsilon^2) \,,
\end{aligned}
\end{equation}
\begin{equation}
    \frac{dE}{d\lambda} = \bra{\Psi(\lambda)}[\hat{H}, \hat{S}(\lambda)]\ket{\Psi(\lambda)} \,,
\end{equation}
\begin{equation}
    \frac{d{}^2D^{i,j}_{k,l}}{d\lambda} = \bra{\Psi(\lambda)}[\hat{\Gamma}^{i,j}_{k,l}, \hat{S}(\lambda)]\ket{\Psi(\lambda)} \,,
\end{equation}
where the operator $\hat{S}$ is defined as:
\begin{equation}
    \hat{S}(\lambda) = \sum_{ijkl} {}^2S^{i,j}_{k,l}\hat{a}^{\dagger}_i \hat{a}^{\dagger}_j \hat{a}_l \hat{a}_k (\lambda)  \,,
\end{equation}
which at each step of $\lambda$ is chosen to minimize the energy along the gradient:
\begin{equation}
    {}^2 S^{i,j}_{k,l}(\lambda) = \bra{\Psi(\lambda)}[\hat{\Gamma}^{i,j}_{k,l},\hat{H}]\ket{\Psi(\lambda)} \,.
\end{equation}
An optimal propagation step in $\lambda$ is determined using a Taylor expansion in approximate form of the 2-RDM
\begin{equation}
    {}^2D(\lambda + \epsilon) \approx {}^2D(\lambda) + \epsilon {}^2D^{'}(\lambda) + \frac{\epsilon^2}{2h} ({}^2D^{'}(\lambda+h) - {}^2D^{'}(\lambda)) \,,
\end{equation}
where h is a small propagation step, and by minimizing the first derivative of the energy, yielding $\epsilon_{\text{opt}}$:
\begin{equation}
    \epsilon_{\text{opt}} = - \frac{\text{Tr}({}^2K{}^2D^{'}(\lambda))}{\text{Tr}({}^2K{}^2D^{''}(\lambda))} \,.
\end{equation}
The 2-RDM ${}^2D$ is updated with the optimal step and this procedure is repeated until a critical value in $\lambda$ is reached where either the energy reaches a minimum or the norm of the residual increases.\\

\section{Discussion and Results}
\subsection{Small Main Group Biradicaloid Set}
\begin{figure}
    \begin{minipage}[b]{0.49\textwidth}
    \centering
    \includegraphics[scale=0.29]{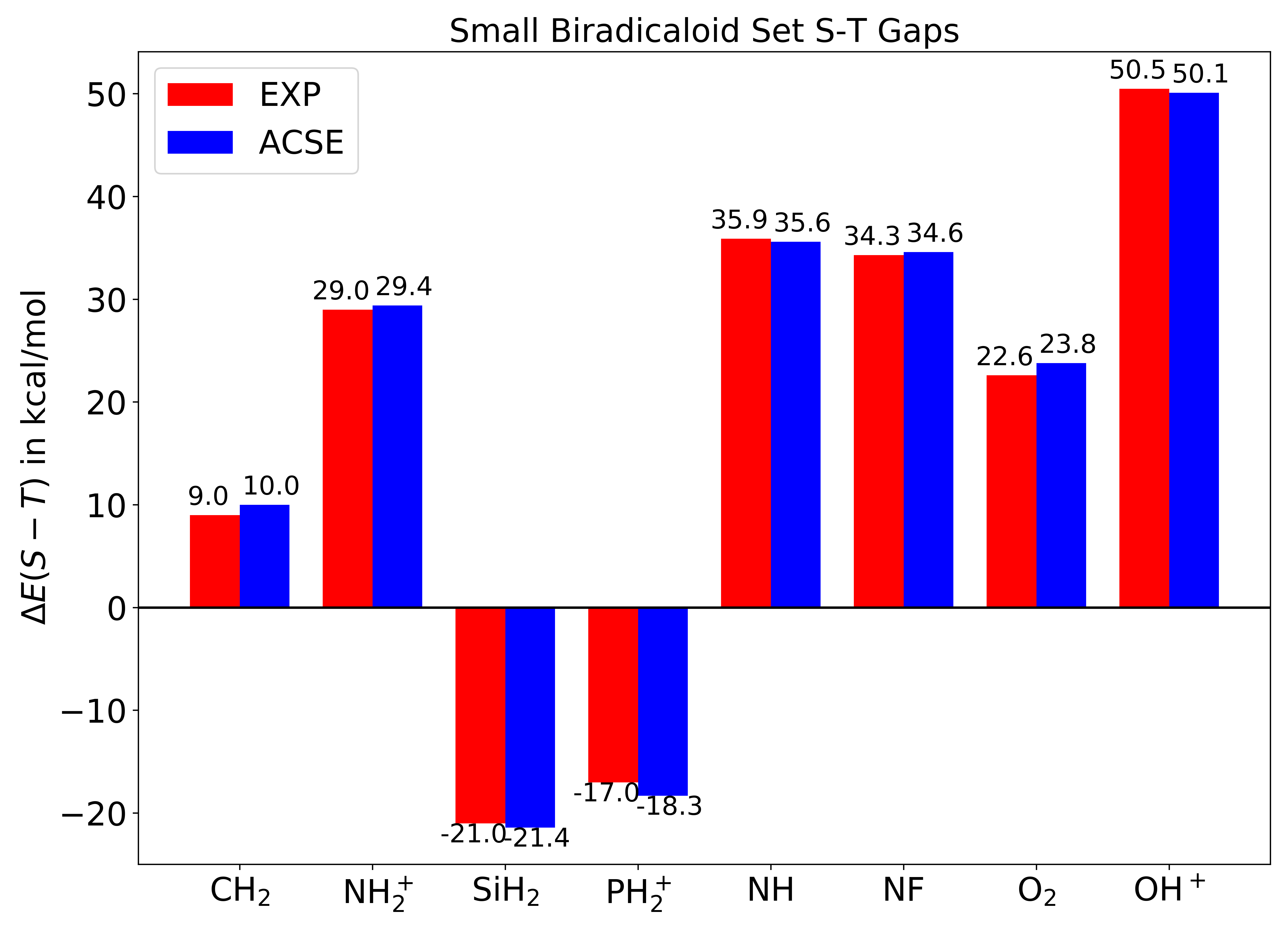}\\
    \end{minipage}
    \begin{minipage}[b]{0.49\textwidth}
    \includegraphics[scale=0.29]{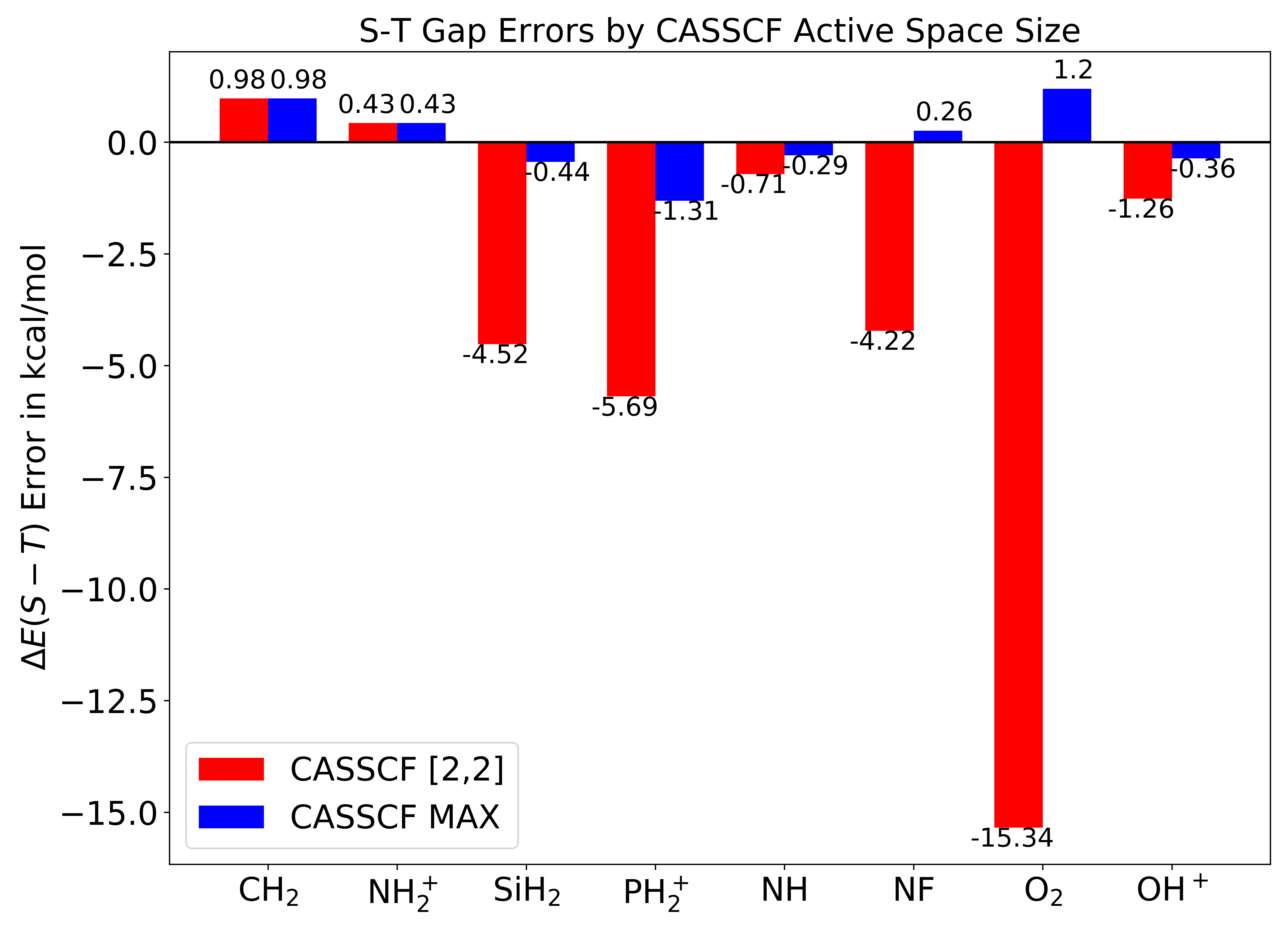}\\
    \end{minipage}
    \caption{Left: Comparison of S-T gaps computed with the converged active-space spin-averaged CASSCF/ACSE method with the cc-pVTZ basis set and experimental values. Experimental results from: OH$^+$, NH, NF, O$_2$ \cite{HuberConst}; NH$_2^+$ \cite{NH2}; CH$_2$ \cite{CH2}; PH$_2^+$\cite{PH2}; SiH$_2$\cite{SiH2}. Right: Errors of converged S-T gaps obtained with a minimal [2,2] active space CASSCF guess and the final converged result seeded with the largest active space guess.}
    \label{fig:ACSEvsExp}
\end{figure}

We first evaluate the spin-averaged ACSE by considering a set of small main-group biradicaloids which have been widely used in the literature to benchmark the accuracy of new computational methods at predicting singlet-triplet gaps in biradical systems\cite{RF-AFQMC, PDFT-MG, KrylovSFDFT, Krylov-SFCCSD, VFS}. The set consists of the 8 species OH$^+$, NH, NF, O$_2$, NH$_2^+$, CH$_2$, PH$_2^+$, SiH$_2$, for all of which experimental S-T gap data is available\cite{HuberConst, NH2, CH2, PH2, SiH2}. Initial calculations were performed with CASSCF in the Maple Quantum Chemistry Package (QCP)\cite{Maple,QCP} for both singlet and triplet spin states using geometries from reference \cite{PDFT-MG} to obtain seed RDMs for the ACSE calculations. All calculations were performed with the cc-pVTZ basis set\cite{ccbasis, ccbasis2}. The left hand side of Figure \ref{fig:ACSEvsExp} shows the results obtained with the CASSCF seeded spin averaged ACSE (CASSCF/ACSE) in comparison to experimental reference values. \\

The spin-averaged ACSE successfully predicts the singlet-triplet gaps in 6 out of the 8 species with chemical accuracy (error of less than 1 kcal/mol) with the largest errors observed for O$_2$ (1.2 kcal/mol) and PH$_2^+$ (1.3 kcal/mol). This large error in O$_2$ coincides with the system demonstrating particularly slow convergence of the S-T gap as a function of the active space size of the initial CASSCF 2-RDM used to seed the calculation with the minimal [2,2] CASSCF guess yielding in an error of -15.34 kcal/mol. The error improves significantly to 1.9 kcal/mol with a moderately large [8,8] active space, finally reducing to 1.3 kcal/mol at  the largest surveyed active space of [14,14]. This error is in general agreement with results obtained by authors of \cite{RF-AFQMC}, which shows errors of 1.7 kcal/mol for active spaces as large as [10,15] in AF-QMC with a CASSCF trial wave function. While O$_2$ proves to be challenging, convergence with respect to CASSCF active space in all other systems is significantly better, and relatively small active spaces are required to obtain accurate results. The right hand side of Figure \ref{fig:ACSEvsExp} compares the deviation from experiment for calculations with a minimal [2,2] CASSCF active space to the final converged result. Both CH$_2$ and NH$_2^+$ show full convergence with a [2,2] guess, while NH and OH$^+$ converge to within 1 kcal/mol of the final result, which is achieved with [4,4] and [6,14] active spaces respectively. SiH, PH$_2^+$ and NF all require larger active spaces for good convergence, however, observed behavior is significantly better than O$_2$, and final results are reported with [10,20] guesses. \textcolor{black}{It is worth noting that generally the ACSE algorithm performs better with a smaller seed active space, where the CI to resolves the multi-reference correlation and the ACSE computes the remainder of the total correlation. Hence, to control the size of the guess RDMs we set an active threshold that excludes orbitals with NON lower than 0.99, limiting the variation in ACSE convergence resulting from the size of the CASSCF RDMs. However, larger active spaces may be needed to resolve the most optimal molecular orbitals or the role of spectator orbitals, giving rise to the varying convergence behavior observed across the different biradicals.} Detailed data on convergence with respect to the size of the CASSCF active space can be found in the Supplemental Information. \\

\begin{table}[h]
    \centering
    \begin{tabular}{c|cc}
        Method & MSE & MAE \\
        \hline
        ACSE & 0.06 & 0.66 \\
        AFQMC/CASSCF\cite{RF-AFQMC} & 0.83 & 1.0 \\
        AFQMC/U\cite{RF-AFQMC} & 0.5 & 1.1 \\
        SF-LDA\cite{KrylovSFDFT} & -0.7 & 1.7 \\
        SF-CIS(D)\cite{Krylov-SFCCSD} & 1.8 & 1.9 \\
        (V)FS-PBE\cite{VFS} & 4.2 & 4.3\\
        W2X\cite{PDFT-MG} & 3 & 3.7 \\
        pp-B3LYP\cite{YangPPrpa} & -3.5 & 4.8 \\
        CASSCF/MC-PDFT\cite{PDFT-MG} & 0.8 & 3.5\\
        CASPT2/MC-PDFT\cite{PDFT-MG} & 0.09 & 0.71\\
        tPBE/MC-PDFT\cite{PDFT-MG} & -4.0 & 4.3 \\
        WABS BLYP\cite{PDFT-MG} & 1.0 & 6.5 \\
    \end{tabular}
    \caption{Comparison of MSE and MAE in the calculation of S-T Gaps in kcal/mol for the benchmarking set consisting of CH$_2$, NH$_2^+$, SiH$_2$, PH$_2^+$, NH, NF, O$_2$, OH$^+$ with ACSE and various methods reported in the literature. ACSE calculations were carried out with a cc-pVTZ basis set.}
    \label{tab:biradicaloids_errors}
\end{table}

The results obtained from the spin-averaged ACSE algorithm compare favorably to other methods reported in the literature, with Table \ref{tab:biradicaloids_errors} showing mean signed errors (MSE) and mean absolute errors (MAE) for the ACSE method and multiple commonly used as well as recently reported methods. The MSE of 0.06 kcal/mol and MAE of 0.66 kcal/mol of this work slightly outperform recently reported AF-QMC/CASSCF (MAE = 0.83 kcal/mol, MSE = 1.0 kcal/mol)\cite{RF-AFQMC} and CASPT2/MC-PDFT (MSE = 0.09 kcal/mol, MAE = 0.71 kcal/mol)\cite{PDFT-MG} values, and provide significantly better results than spin flip or particle-particle random phase approximation based methods.

\subsection{Small Organic Biradicals}
Not only have organic biradical and multiradical compounds attracted significant research interest in the design of molecular magnets, conductors, and the field of spintronics\cite{FeTTfttdirad, OrgBirad, OrgBirad2, OrgBirad3, OrgBirad4}, they also play important roles in chemical reactions as transition states or reactive intermediates, making their accurate depiction essential for the computational modelling and elucidation of chemical processes\cite{birad,Birad1, Birad2, Birad3, carbenes, NCarb}. Nevertheless, their exact treatment remains a challenge in contemporary electronic structure theory. In this section we investigate several representative small organic biradical systems that have been studied extensively in the literature. This includes the ortho-, meta-, and para- isomers of benzyne, derivatives of which have gained traction in recent years as important intermediates in organic synthesis\cite{BenzynesSynth1, BenzyneSynth2, BenzyneSynth3, mpBenzyneRev}, as well as the small four electron biradicals trimethylenemethane (TMM) and cyclobutadiene. \\

While the three benzyne isomers exhibit singlet ground states, they have a strongly correlated electronic structure arising from their biradical nature. The magnitude of their T-S gap follows an inverse relationship with the degree to which the isomer exhibits diradical character. Experimental data have been reported for the gas phase, with the para isomer exhibiting the smallest adiabatic gap of 3.3 kcal/mol , followed by meta (20.0 kcal/mol) and ortho (38.1 kcal/mol)\cite{BenzyneExp}. Our ACSE calculations employ the cc-pVDZ basis set in conjunction with a [4,4] active space. Geometries were obtained from reference \cite{YangPPrpa}, producing T-S gaps of 35.79 kcal/mol, 20.66 kcal/mol and 7.49 kcal/mol for the ortho, meta and para isomers respectively. In the cases of all isomers, no gain in accuracy was achieved when using a larger active space to generate the CASSCF 2-RDM used to seed the ACSE calculation. Increasing the active space from the minimal [2,2] space, which only includes the radical orbitals, to the [4,4] space gave a slight improvement, while any further increases leads to inclusion of molecular orbitals in the active space that show no significant correlation and exhibit occupation numbers of  > 0.99 or < 0.01, resulting in divergence away from the experimental results. This is in line with previous observations on the ACSE and its dependence on the initial guess\cite{ACSEcor,ACSEextrap}, showing that the optimal CASSCF guess should only account for significantly correlated space, allowing the ACSE algorithm to resolve the entirety of the remaining correlation. \\

\begin{table}[]
    \centering
    \begin{tabular}{c|ccc}
         & \multicolumn{3}{c}{$\Delta E_{T-S}$} \\
         & o- & m- & p- \\
         \hline
        Exp\cite{BenzyneExp} & 38.1 & 20.0 & 3.3 \\
        ACSE & 35.79 & 20.66 & 7.49  \\
        AFQMC/CAS\cite{RF-AFQMC} & 37.4(6) & 20.7(8) & 4.5(5) \\
        AFQMC/U\cite{RF-AFQMC} & 37.6(7) & 18.9(9) & 2.2(9) \\
        pp-HF\cite{YangPPrpa} & 45.6 & 35.5 & 4.0 \\
        pp-B3LYP\cite{YangPPrpa}  & 37.4 & 22.1 & 0.6 \\
        SF-CIS(D)\cite{Krylov-SFCCSD} & 35.7 & 19.4 & 2.1 \\
        SF-CCSD(T)\cite{Krylov-SF-CCSDT} & 37.3 & 20.6 & 4.0 \\
        SF-LDA\cite{KrylovSFDFT} & 47.5 & 29.0 & 10.5 \\
        UB3LYP\cite{UB3LYP} & 29.4 & 14.2 & 2.4 \\
    \end{tabular}
    \caption{S-T gaps in kcal/mol for the ortho-, meta- and para-benzyne isomers obtained with ACSE and the cc-pVDZ basis set, compared to ZPE corrected experimental gaps and those obtained via several methods reported in the literature.}
    \label{tab:benzynes}
\end{table}

Our results are compared with a range of contemporary methods in Table \ref{tab:benzynes}. The ACSE provides good agreement with experiment in the case of the meta isomer, comparing favorably to the other surveyed methods, and yielding identical results to AFQMC/CAS. The results obtained for the ortho isomer are also satisfactory, being comparable to SF-CIS(D) but producing slightly larger errors than AFQMC/CAS and SF-CCSD(T), which yield gaps of 37.4 kcal/mol and 37.3 kcal/mol respectively. Finally, in the case of the para isomer the ACSE calculation yields an error of 4.2 kcal/mol, which, while small, only outperforms SF-LDA, which displays an error of 7.2 kcal/mol. The larger error in the para isomer coincides with the greatest contribution of strong correlation to its electronic structure, with the CASSCF reference giving frontier NON for the HONO and LUNO of $\lambda_{\text{HONO}} = 0.610 $ and $\lambda_{\text{LUMO}} =  0.390 $. The significant multi-reference character results in a particularly large stabilization of the singlet state with respect to the triplet, which in turn is not fully compensated by the additional dynamic correlation recovered by the ACSE, leading to an overestimation the gap. Additionally, a residual source of error may be the use of the relatively small cc-pVDZ basis set when compared to experiment and calculations using the larger cc-pVTZ and cc-pVQZ basis sets. \\

In addition to the benzynes we investigate the two small four-electron biradicals TMM and cyclobutadiene. Geometries were obtained from reference \cite{YangPPrpa} and the calculations were carried out with the cc-pVDZ basis set. TMM is a highly reactive radical and the simplest molecule that cannot be described by a Kekule structure\cite{TMM}. Resulting from the four carbon p-orbitals, four $\pi$-orbitals split into one filled orbital, two degenerate orbitals occupied by two electrons, and one unfilled orbital. Obeying Hund's rule, its electronic ground state is an open shell triplet with D$_{3h}$ symmetry, while the first low lying excited state is a singlet that breaks the D$_{3h}$ symmetry and instead assumes a twisted non-planar geometry. The ACSE, seeded with a [4,4] CASSCF guess, predicts a S-T gap of 16.60 kcal/mol, which closely mirrors the results obtained from MCQDPT2 [10,10] (16.7 kcal/mol) and EOM-CCSD(dT) (15.3 kcal/mol). The results are summarized in the first row of Table \ref{tab:small_orgs}. \\

\begin{table*}[]
    \centering
    \resizebox{\textwidth}{!}{
    \begin{tabular}{cc|cccccccccc}
       & & MCQDPT2 & & & pp- & pp- & (V)FS- & SF- & SF- & DEA- & CASSCF-\\
        & & [10,10] & ACSE & (dT)\cite{Krylov-SF-CCSDT} & HF\cite{YangPPrpa} & B3LYP\cite{YangPPrpa} & PBE\cite{VFS} & CIS(D)\cite{Krylov-SFCCSD} & LDA\cite{KrylovSFDFT} & EOMCC\cite{Laura_MRCC} & MkCCSD\cite{mkCCSD}\\
        \hline
        \multirow{2}{*}{$\Delta E_{T-S}$} & TMM & 16.7 & 16.60 & 16 & 17.1 & 16.3 & 21.5 & 23.6 & 15.6 & &  \\
        & cyclobutadiene & & -7.46 & 7.2 & 6.7 & 6.5 & & & & -4.2 & -8.1\\
    \end{tabular}}
    \caption{S-T gaps in kcal/mol calculated with the ACSE method with a cc-pVDZ basis set for cyclobutadiene and TMM compared with gaps reported in the literature. }
    \label{tab:small_orgs}
\end{table*}

Analogous to TMM cyclobutadiene has four carbon p-orbitals forming four $\pi$-orbitals occupied by four electrons. However, while the orbital picture is identical to TMM, the splitting of the spin states is not --- in cyclobutadiene the two unpaired $\pi$ electrons form a singlet ground state with a triplet excited state, violating Hund's rule. Here we investigate the D$_4h$ square symmetric geometry of the molecule, which has previously been theoretically investigated by various methods. It is worth noting that experimentally it has been observed that the singlet is not stable in the D$_4h$ geometry due to Jahn-Teller distortion and instead cyclobutadiene adopts a rectangular structure\cite{cyclobutadiene}. Using a [12,12] CASSCF guess, the ACSE predicts a triplet ground state with a S-T gap of -7.46 kcal/mol. This result agrees with reference data obtained at the DEA-EOMCC and CASSCF-MkCCSD levels of theory, which yield gaps of -4.2 kcal/mol and -8.1 kcal/mol respectively, while pp-RPA based methods and EOM-CCSD(dT) fail to predict the correct ordering of singlet and triplet states (see second row of Table \ref{tab:small_orgs}). Cyclobutadiene presents a system that benefits from the use of an increased active space, with a [4,4] CASSCF guess producing a S-T gap of -11.33 kcal/mol. This, however, is driven by the role of spectator orbitals in the CASSCF calculation. The [4,4] active space produces degenerate HONO and LUNO populations of $\lambda = 0.5$, while the [12,12] CASSCF converges to non-degenerate populations of $\lambda_{\text{HONO}} = 0.65 $ and $\lambda_{\text{LUMO}} =  0.35$, indicating the variation in results can be attributed to major changes in the static correlation across the active spaces in the CASSCF calculation of the singlet state. \\

\subsection{Naphthalene}
Lastly, we investigate the ability of the spin averaged ACSE method to resolve the S-T gaps in polyacenes. Polyacenes are the building blocks of the narrowest graphene nanoribbons, and due to their tunable electronic properties they are widely studied for their roles in potential semiconductors or field effect transistors\cite{Acenes1, Acenes2}. Recent experiments with acene chains up to decacene have shown finite band gaps in the large polymer limit\cite{aceneexp}, however, electronic structure methods that fail to account for strong correlation such as DFT predict a crossing of the spin states towards a singlet ground state with increased system size at finite ring size\cite{acene-TDDFT}. As such acenes remain a prime example for a molecular system that necessitates methods that accurately account for both static and dynamic correlation to hold value as a predictive computational tool. In this section we investigate naphthalene, the smallest member of the polyacene family, to evaluate the ACSE's ability to resolve qualities of the electronic structure of larger organic systems. \\

\begin{table}[]
    \centering
    \begin{tabular}{c|c}
         & $\Delta E_{T-S}$ \\
         \hline
        Exp\cite{Naphexp,Naphzpe} & 64.4 \\
        CASSCF & 65.78\\
        ACSE & 67.86 \\
        GAS-PDFT\cite{LauraDMRG} &  70.6 \\
        DMRG\cite{LauraDMRG} & 61.3 \\
        DMRG-PDFT\cite{LauraDMRG} & 67.1 \\
        DMRG-ec-MRCISD+Q\cite{LauraDMRG} & 62.5 \\
        AFQMC/U\cite{RF-AFQMC} & 68 (1.2) \\
        CCSD(T)/FPA\cite{FPACCSDT} & 65.8 \\
        pp-B3LYP\cite{YangPPrpa} & 66.2 \\
    \end{tabular}
    \caption{T-S gap in kcal/mol obtained with CASSCF/ACSE using a [12,12] active space and cc-pVDZ basis set, compared with ZPE corrected experimental value, and a range of data points reported in the literature.}
    \label{tab:naph}
\end{table}

Geometries for the singlet and triplet states were obtained from reference \cite{NaphGeo} and the ACSE calculations were seeded with a 2-RDM obtained from a [12,12] CASSCF calculation. The cc-pVDZ basis set was used for all calculations and the results are summarized and compared to values reported in the literature in Table \ref{tab:naph}. The experimental T-S gap of 64.4 kcal/mol denotes the zero-point energy corrected value, based on the experimentally measured result of 61.0 kcal/mol minus the adiabatic correction energy of -3.4 kcal/mol as reported for B3LYP/6-31G(d)\cite{Naphzpe}.  It is worth noting that experimental data for larger, stable molecules such as naphthalene is obtained under real world laboratory conditions and as such comparison of ab-initio calculations assuming a zero temperature, gas phase environment will always suffer from residual errors. The ACSE value of 67.86 kcal/mol compares favorably to other state of the art methods reported in the literature such as DMRG, PDFT based methods or AFQMC, for which in this case only results seeded with a UDFT guess are available. Overestimation of the gap is in line with the rest of the surveyed methods, with five out of seven yielding a positive deviation from experiment. The spin-free CASSCF/ACSE yields results that show promise for its application in strongly correlated, closed shell organic molecules.  \\

\section{Conclusions}
We have developed a spin-averaging modification to the ACSE, significantly reducing the complexity of state and input preparation, which required coupling to H atoms to yield singlet systems in previous implementations. Using the spin averaging scheme detailed in this article, we can treat open-shell singlets and higher multiplicity states in the same way as closed shell ground states and without additional file or input preparation. To demonstrate the ability of the spin-averaged ACSE to resolve both dynamic and strong correlation, we successfully calculated the singlet-triplet gaps in a small main group biradicaloid benchmark set, as well as several small organic molecules. We show that the spin-averaged ACSE yields results of comparable accuracy to other state-of-the-art methods such as pp-RPA or AFQMC or PDFT reported in the literature. This is a major step in the development of the ACSE into an easily applied computational method to resolve the electronic structure of molecules that require the resolution of both strong and dynamic correlation for an accurate treatment. \\

Future work is being undertaken to use the spin-averaged ACSE to resolve not only ground states but also excited states of a specified multiplicity, allowing it to resolve spectral information. We also aim to extend the spin-averaged ACSE to include odd-electron count open-shell systems, with applications the spin state splittings and magnetic coupling in transition metal complexes. The latter pose particularly great and relevant challenges to modern quantum chemistry, as they are oftentimes being distinguished by multi-reference ground states with significant dynamical correlation, while being particularly pertinent to large areas of modern synthetic chemistry and materials science\cite{CoDimer, LauraDimer}.

\noindent {\bf Data Availability Statement:} The data that support the findings of this study are available from the corresponding author upon reasonable request.

\noindent {\bf Supplementary Material:} The supplementary material includes the active-space-size dependence of the CASSCF/ACSE  singlet-triplet-gap calculations of the main group biradicaloids.

\begin{acknowledgments}
D.A.M. gratefully acknowledges the U.S. National Science Foundation Grant No. CHE-1565638.
\end{acknowledgments}

\bibliographystyle{apsrev4-1}
\bibliography{citations.bib}

\begin{thebibliography}{89}%
\makeatletter
\providecommand \@ifxundefined [1]{%
 \@ifx{#1\undefined}
}%
\providecommand \@ifnum [1]{%
 \ifnum #1\expandafter \@firstoftwo
 \else \expandafter \@secondoftwo
 \fi
}%
\providecommand \@ifx [1]{%
 \ifx #1\expandafter \@firstoftwo
 \else \expandafter \@secondoftwo
 \fi
}%
\providecommand \natexlab [1]{#1}%
\providecommand \enquote  [1]{``#1''}%
\providecommand \bibnamefont  [1]{#1}%
\providecommand \bibfnamefont [1]{#1}%
\providecommand \citenamefont [1]{#1}%
\providecommand \href@noop [0]{\@secondoftwo}%
\providecommand \href [0]{\begingroup \@sanitize@url \@href}%
\providecommand \@href[1]{\@@startlink{#1}\@@href}%
\providecommand \@@href[1]{\endgroup#1\@@endlink}%
\providecommand \@sanitize@url [0]{\catcode `\\12\catcode `\$12\catcode
  `\&12\catcode `\#12\catcode `\^12\catcode `\_12\catcode `\%12\relax}%
\providecommand \@@startlink[1]{}%
\providecommand \@@endlink[0]{}%
\providecommand \url  [0]{\begingroup\@sanitize@url \@url }%
\providecommand \@url [1]{\endgroup\@href {#1}{\urlprefix }}%
\providecommand \urlprefix  [0]{URL }%
\providecommand \Eprint [0]{\href }%
\providecommand \doibase [0]{http://dx.doi.org/}%
\providecommand \selectlanguage [0]{\@gobble}%
\providecommand \bibinfo  [0]{\@secondoftwo}%
\providecommand \bibfield  [0]{\@secondoftwo}%
\providecommand \translation [1]{[#1]}%
\providecommand \BibitemOpen [0]{}%
\providecommand \bibitemStop [0]{}%
\providecommand \bibitemNoStop [0]{.\EOS\space}%
\providecommand \EOS [0]{\spacefactor3000\relax}%
\providecommand \BibitemShut  [1]{\csname bibitem#1\endcsname}%
\let\auto@bib@innerbib\@empty
%</preamble>
\bibitem [{\citenamefont {Scheschkewitz}\ \emph {et~al.}(2002)\citenamefont
  {Scheschkewitz}, \citenamefont {Amii}, \citenamefont {Gornitzka},
  \citenamefont {Schoeller}, \citenamefont {Bourissou},\ and\ \citenamefont
  {Bertrand}}]{birad}%
  \BibitemOpen
  \bibfield  {author} {\bibinfo {author} {\bibfnamefont {D.}~\bibnamefont
  {Scheschkewitz}}, \bibinfo {author} {\bibfnamefont {H.}~\bibnamefont {Amii}},
  \bibinfo {author} {\bibfnamefont {H.}~\bibnamefont {Gornitzka}}, \bibinfo
  {author} {\bibfnamefont {W.~W.}\ \bibnamefont {Schoeller}}, \bibinfo {author}
  {\bibfnamefont {D.}~\bibnamefont {Bourissou}}, \ and\ \bibinfo {author}
  {\bibfnamefont {G.}~\bibnamefont {Bertrand}},\ }\href@noop {} {\bibfield
  {journal} {\bibinfo  {journal} {Science}\ }\textbf {\bibinfo {volume}
  {295}},\ \bibinfo {pages} {1880} (\bibinfo {year} {2002})}\BibitemShut
  {NoStop}%
\bibitem [{\citenamefont {Hatzimarinaki}\ \emph {et~al.}(2005)\citenamefont
  {Hatzimarinaki}, \citenamefont {Roubelakis},\ and\ \citenamefont
  {Orfanopoulos}}]{Rxn1}%
  \BibitemOpen
  \bibfield  {author} {\bibinfo {author} {\bibfnamefont {M.}~\bibnamefont
  {Hatzimarinaki}}, \bibinfo {author} {\bibfnamefont {M.~M.}\ \bibnamefont
  {Roubelakis}}, \ and\ \bibinfo {author} {\bibfnamefont {M.}~\bibnamefont
  {Orfanopoulos}},\ }\href@noop {} {\bibfield  {journal} {\bibinfo  {journal}
  {J. Am. Chem. Soc.}\ }\textbf {\bibinfo {volume} {127}},\ \bibinfo {pages}
  {14182} (\bibinfo {year} {2005})}\BibitemShut {NoStop}%
\bibitem [{\citenamefont {Snyder}(1989)}]{Rxn2}%
  \BibitemOpen
  \bibfield  {author} {\bibinfo {author} {\bibfnamefont {J.~P.}\ \bibnamefont
  {Snyder}},\ }\href@noop {} {\bibfield  {journal} {\bibinfo  {journal} {J. Am.
  Chem. Soc.}\ }\textbf {\bibinfo {volume} {111}},\ \bibinfo {pages} {7630}
  (\bibinfo {year} {1989})}\BibitemShut {NoStop}%
\bibitem [{\citenamefont {O'Neal}\ and\ \citenamefont {Benson}(1968)}]{Rxn3}%
  \BibitemOpen
  \bibfield  {author} {\bibinfo {author} {\bibfnamefont {H.~E.}\ \bibnamefont
  {O'Neal}}\ and\ \bibinfo {author} {\bibfnamefont {S.~W.}\ \bibnamefont
  {Benson}},\ }\href@noop {} {\bibfield  {journal} {\bibinfo  {journal} {J.
  Phys. Chem.}\ }\textbf {\bibinfo {volume} {72}},\ \bibinfo {pages} {1866}
  (\bibinfo {year} {1968})}\BibitemShut {NoStop}%
\bibitem [{\citenamefont {Doubleday}\ \emph {et~al.}(1988)\citenamefont
  {Doubleday}, \citenamefont {McIver},\ and\ \citenamefont {Page}}]{Rxn4}%
  \BibitemOpen
  \bibfield  {author} {\bibinfo {author} {\bibfnamefont {C.}~\bibnamefont
  {Doubleday}}, \bibinfo {author} {\bibfnamefont {J.~W.}\ \bibnamefont
  {McIver}}, \ and\ \bibinfo {author} {\bibfnamefont {M.}~\bibnamefont
  {Page}},\ }\href@noop {} {\bibfield  {journal} {\bibinfo  {journal} {J. Phys.
  Chem.}\ }\textbf {\bibinfo {volume} {92}},\ \bibinfo {pages} {4367} (\bibinfo
  {year} {1988})}\BibitemShut {NoStop}%
\bibitem [{\citenamefont {Bourissou}\ \emph
  {et~al.}(2000{\natexlab{a}})\citenamefont {Bourissou}, \citenamefont
  {Guerret}, \citenamefont {Gabbaï},\ and\ \citenamefont
  {Bertrand}}]{carbenes}%
  \BibitemOpen
  \bibfield  {author} {\bibinfo {author} {\bibfnamefont {D.}~\bibnamefont
  {Bourissou}}, \bibinfo {author} {\bibfnamefont {O.}~\bibnamefont {Guerret}},
  \bibinfo {author} {\bibfnamefont {F.~P.}\ \bibnamefont {Gabbaï}}, \ and\
  \bibinfo {author} {\bibfnamefont {G.}~\bibnamefont {Bertrand}},\ }\href@noop
  {} {\bibfield  {journal} {\bibinfo  {journal} {Chem. Rev.}\ }\textbf
  {\bibinfo {volume} {100}},\ \bibinfo {pages} {39} (\bibinfo {year}
  {2000}{\natexlab{a}})}\BibitemShut {NoStop}%
\bibitem [{\citenamefont {Bourissou}\ \emph
  {et~al.}(2000{\natexlab{b}})\citenamefont {Bourissou}, \citenamefont
  {Guerret}, \citenamefont {Gabba{\"i}},\ and\ \citenamefont
  {Bertrand}}]{NCarb}%
  \BibitemOpen
  \bibfield  {author} {\bibinfo {author} {\bibfnamefont {D.}~\bibnamefont
  {Bourissou}}, \bibinfo {author} {\bibfnamefont {O.}~\bibnamefont {Guerret}},
  \bibinfo {author} {\bibfnamefont {F.~P.}\ \bibnamefont {Gabba{\"i}}}, \ and\
  \bibinfo {author} {\bibfnamefont {G.}~\bibnamefont {Bertrand}},\ }\href@noop
  {} {\bibfield  {journal} {\bibinfo  {journal} {Chem. Rev.}\ }\textbf
  {\bibinfo {volume} {100}},\ \bibinfo {pages} {39} (\bibinfo {year}
  {2000}{\natexlab{b}})}\BibitemShut {NoStop}%
\bibitem [{\citenamefont {Stuyver}\ \emph {et~al.}(2019)\citenamefont
  {Stuyver}, \citenamefont {Chen}, \citenamefont {Zeng}, \citenamefont
  {Geerlings}, \citenamefont {De~Proft},\ and\ \citenamefont
  {Hoffmann}}]{Birad1}%
  \BibitemOpen
  \bibfield  {author} {\bibinfo {author} {\bibfnamefont {T.}~\bibnamefont
  {Stuyver}}, \bibinfo {author} {\bibfnamefont {B.}~\bibnamefont {Chen}},
  \bibinfo {author} {\bibfnamefont {T.}~\bibnamefont {Zeng}}, \bibinfo {author}
  {\bibfnamefont {P.}~\bibnamefont {Geerlings}}, \bibinfo {author}
  {\bibfnamefont {F.}~\bibnamefont {De~Proft}}, \ and\ \bibinfo {author}
  {\bibfnamefont {R.}~\bibnamefont {Hoffmann}},\ }\href@noop {} {\bibfield
  {journal} {\bibinfo  {journal} {Chem. Rev.}\ }\textbf {\bibinfo {volume}
  {119}},\ \bibinfo {pages} {11291} (\bibinfo {year} {2019})}\BibitemShut
  {NoStop}%
\bibitem [{\citenamefont {Salem}\ and\ \citenamefont {Rowland}(1972)}]{Birad2}%
  \BibitemOpen
  \bibfield  {author} {\bibinfo {author} {\bibfnamefont {L.}~\bibnamefont
  {Salem}}\ and\ \bibinfo {author} {\bibfnamefont {C.}~\bibnamefont
  {Rowland}},\ }\href@noop {} {\bibfield  {journal} {\bibinfo  {journal}
  {Angew. Chem.}\ }\textbf {\bibinfo {volume} {11}},\ \bibinfo {pages} {92}
  (\bibinfo {year} {1972})}\BibitemShut {NoStop}%
\bibitem [{\citenamefont {Abe}(2013)}]{Birad3}%
  \BibitemOpen
  \bibfield  {author} {\bibinfo {author} {\bibfnamefont {M.}~\bibnamefont
  {Abe}},\ }\href@noop {} {\bibfield  {journal} {\bibinfo  {journal} {Chem.
  Rev.}\ }\textbf {\bibinfo {volume} {113}},\ \bibinfo {pages} {7011} (\bibinfo
  {year} {2013})}\BibitemShut {NoStop}%
\bibitem [{\citenamefont {Burke}(2012)}]{DFT1}%
  \BibitemOpen
  \bibfield  {author} {\bibinfo {author} {\bibfnamefont {K.}~\bibnamefont
  {Burke}},\ }\href@noop {} {\bibfield  {journal} {\bibinfo  {journal} {J.
  Chem. Phys.}\ }\textbf {\bibinfo {volume} {136}},\ \bibinfo {pages} {150901}
  (\bibinfo {year} {2012})}\BibitemShut {NoStop}%
\bibitem [{\citenamefont {Jain}\ \emph {et~al.}(2016)\citenamefont {Jain},
  \citenamefont {Shin},\ and\ \citenamefont {Persson}}]{DFT2}%
  \BibitemOpen
  \bibfield  {author} {\bibinfo {author} {\bibfnamefont {A.}~\bibnamefont
  {Jain}}, \bibinfo {author} {\bibfnamefont {Y.}~\bibnamefont {Shin}}, \ and\
  \bibinfo {author} {\bibfnamefont {K.~A.}\ \bibnamefont {Persson}},\
  }\href@noop {} {\bibfield  {journal} {\bibinfo  {journal} {Nat. Rev. Mater.}\
  }\textbf {\bibinfo {volume} {1}},\ \bibinfo {pages} {15004} (\bibinfo {year}
  {2016})}\BibitemShut {NoStop}%
\bibitem [{\citenamefont {Grimme}\ and\ \citenamefont
  {Schreiner}(2018)}]{challenges}%
  \BibitemOpen
  \bibfield  {author} {\bibinfo {author} {\bibfnamefont {S.}~\bibnamefont
  {Grimme}}\ and\ \bibinfo {author} {\bibfnamefont {P.~R.}\ \bibnamefont
  {Schreiner}},\ }\href@noop {} {\bibfield  {journal} {\bibinfo  {journal}
  {Angew. Chem.}\ }\textbf {\bibinfo {volume} {57}},\ \bibinfo {pages} {4170}
  (\bibinfo {year} {2018})}\BibitemShut {NoStop}%
\bibitem [{\citenamefont {Perdew}\ \emph {et~al.}(2009)\citenamefont {Perdew},
  \citenamefont {Ruzsinszky}, \citenamefont {Constantin}, \citenamefont {Sun},\
  and\ \citenamefont {Csonka}}]{DFTfundamentalissues}%
  \BibitemOpen
  \bibfield  {author} {\bibinfo {author} {\bibfnamefont {J.~P.}\ \bibnamefont
  {Perdew}}, \bibinfo {author} {\bibfnamefont {A.}~\bibnamefont {Ruzsinszky}},
  \bibinfo {author} {\bibfnamefont {L.~A.}\ \bibnamefont {Constantin}},
  \bibinfo {author} {\bibfnamefont {J.}~\bibnamefont {Sun}}, \ and\ \bibinfo
  {author} {\bibfnamefont {G.~I.}\ \bibnamefont {Csonka}},\ }\href@noop {}
  {\bibfield  {journal} {\bibinfo  {journal} {J. Chem. Theory Comput.}\
  }\textbf {\bibinfo {volume} {5}},\ \bibinfo {pages} {902} (\bibinfo {year}
  {2009})}\BibitemShut {NoStop}%
\bibitem [{\citenamefont {Raghavachari}\ and\ \citenamefont
  {Anderson}(1996)}]{correlation}%
  \BibitemOpen
  \bibfield  {author} {\bibinfo {author} {\bibfnamefont {K.}~\bibnamefont
  {Raghavachari}}\ and\ \bibinfo {author} {\bibfnamefont {J.~B.}\ \bibnamefont
  {Anderson}},\ }\href@noop {} {\bibfield  {journal} {\bibinfo  {journal} {J.
  Phys. Chem.}\ }\textbf {\bibinfo {volume} {100}},\ \bibinfo {pages} {12960}
  (\bibinfo {year} {1996})}\BibitemShut {NoStop}%
\bibitem [{\citenamefont {Benavides-Riveros}\ \emph {et~al.}(2017)\citenamefont
  {Benavides-Riveros}, \citenamefont {Lathiotakis},\ and\ \citenamefont
  {Marques}}]{correlation2}%
  \BibitemOpen
  \bibfield  {author} {\bibinfo {author} {\bibfnamefont {C.~L.}\ \bibnamefont
  {Benavides-Riveros}}, \bibinfo {author} {\bibfnamefont {N.~N.}\ \bibnamefont
  {Lathiotakis}}, \ and\ \bibinfo {author} {\bibfnamefont {M.~A.~L.}\
  \bibnamefont {Marques}},\ }\href@noop {} {\bibfield  {journal} {\bibinfo
  {journal} {Phys. Chem. Chem. Phys.}\ }\textbf {\bibinfo {volume} {19}},\
  \bibinfo {pages} {12655} (\bibinfo {year} {2017})}\BibitemShut {NoStop}%
\bibitem [{\citenamefont {Cohen}\ \emph {et~al.}(2008)\citenamefont {Cohen},
  \citenamefont {Mori-Sánchez},\ and\ \citenamefont {Yang}}]{FSDFTerror}%
  \BibitemOpen
  \bibfield  {author} {\bibinfo {author} {\bibfnamefont {A.~J.}\ \bibnamefont
  {Cohen}}, \bibinfo {author} {\bibfnamefont {P.}~\bibnamefont
  {Mori-Sánchez}}, \ and\ \bibinfo {author} {\bibfnamefont {W.}~\bibnamefont
  {Yang}},\ }\href@noop {} {\bibfield  {journal} {\bibinfo  {journal} {J. Chem.
  Phys.}\ }\textbf {\bibinfo {volume} {129}},\ \bibinfo {pages} {121104}
  (\bibinfo {year} {2008})}\BibitemShut {NoStop}%
\bibitem [{\citenamefont {Olsen}(2011)}]{CASSCF}%
  \BibitemOpen
  \bibfield  {author} {\bibinfo {author} {\bibfnamefont {J.}~\bibnamefont
  {Olsen}},\ }\href@noop {} {\bibfield  {journal} {\bibinfo  {journal} {Int. J.
  Quantum Chem.}\ }\textbf {\bibinfo {volume} {111}},\ \bibinfo {pages} {3267}
  (\bibinfo {year} {2011})}\BibitemShut {NoStop}%
\bibitem [{\citenamefont {Gräfenstein}\ and\ \citenamefont
  {Cremer}(2000)}]{DFTbirad1}%
  \BibitemOpen
  \bibfield  {author} {\bibinfo {author} {\bibfnamefont {J.}~\bibnamefont
  {Gräfenstein}}\ and\ \bibinfo {author} {\bibfnamefont {D.}~\bibnamefont
  {Cremer}},\ }\href@noop {} {\bibfield  {journal} {\bibinfo  {journal} {Phys.
  Chem. Chem. Phys.}\ }\textbf {\bibinfo {volume} {2}},\ \bibinfo {pages}
  {2091} (\bibinfo {year} {2000})}\BibitemShut {NoStop}%
\bibitem [{\citenamefont {Gräfenstein}\ \emph {et~al.}(2002)\citenamefont
  {Gräfenstein}, \citenamefont {Kraka}, \citenamefont {Filatov},\ and\
  \citenamefont {Cremer}}]{UDFT}%
  \BibitemOpen
  \bibfield  {author} {\bibinfo {author} {\bibfnamefont {J.}~\bibnamefont
  {Gräfenstein}}, \bibinfo {author} {\bibfnamefont {E.}~\bibnamefont {Kraka}},
  \bibinfo {author} {\bibfnamefont {M.}~\bibnamefont {Filatov}}, \ and\
  \bibinfo {author} {\bibfnamefont {D.}~\bibnamefont {Cremer}},\ }\href@noop {}
  {\bibfield  {journal} {\bibinfo  {journal} {Int. J. Mol. Sci.}\ }\textbf
  {\bibinfo {volume} {3}},\ \bibinfo {pages} {360} (\bibinfo {year}
  {2002})}\BibitemShut {NoStop}%
\bibitem [{\citenamefont {Baker}\ \emph {et~al.}(1993)\citenamefont {Baker},
  \citenamefont {Scheiner},\ and\ \citenamefont {Andzelm}}]{UDFT2}%
  \BibitemOpen
  \bibfield  {author} {\bibinfo {author} {\bibfnamefont {J.}~\bibnamefont
  {Baker}}, \bibinfo {author} {\bibfnamefont {A.}~\bibnamefont {Scheiner}}, \
  and\ \bibinfo {author} {\bibfnamefont {J.}~\bibnamefont {Andzelm}},\
  }\href@noop {} {\bibfield  {journal} {\bibinfo  {journal} {Chem. Phys.
  Lett.}\ }\textbf {\bibinfo {volume} {216}},\ \bibinfo {pages} {380 }
  (\bibinfo {year} {1993})}\BibitemShut {NoStop}%
\bibitem [{\citenamefont {Yang}\ \emph {et~al.}(2015)\citenamefont {Yang},
  \citenamefont {Peng}, \citenamefont {Davidson},\ and\ \citenamefont
  {Yang}}]{YangPPrpa}%
  \BibitemOpen
  \bibfield  {author} {\bibinfo {author} {\bibfnamefont {Y.}~\bibnamefont
  {Yang}}, \bibinfo {author} {\bibfnamefont {D.}~\bibnamefont {Peng}}, \bibinfo
  {author} {\bibfnamefont {E.~R.}\ \bibnamefont {Davidson}}, \ and\ \bibinfo
  {author} {\bibfnamefont {W.}~\bibnamefont {Yang}},\ }\href@noop {} {\bibfield
   {journal} {\bibinfo  {journal} {J. Phys. Chem. A}\ }\textbf {\bibinfo
  {volume} {119}},\ \bibinfo {pages} {4923} (\bibinfo {year}
  {2015})}\BibitemShut {NoStop}%
\bibitem [{\citenamefont {van Aggelen}\ \emph {et~al.}(2013)\citenamefont {van
  Aggelen}, \citenamefont {Yang},\ and\ \citenamefont {Yang}}]{pprpa1}%
  \BibitemOpen
  \bibfield  {author} {\bibinfo {author} {\bibfnamefont {H.}~\bibnamefont {van
  Aggelen}}, \bibinfo {author} {\bibfnamefont {Y.}~\bibnamefont {Yang}}, \ and\
  \bibinfo {author} {\bibfnamefont {W.}~\bibnamefont {Yang}},\ }\href@noop {}
  {\bibfield  {journal} {\bibinfo  {journal} {Phys. Rev. A}\ }\textbf {\bibinfo
  {volume} {88}},\ \bibinfo {pages} {030501} (\bibinfo {year}
  {2013})}\BibitemShut {NoStop}%
\bibitem [{\citenamefont {Yang}\ \emph {et~al.}(2013)\citenamefont {Yang},
  \citenamefont {van Aggelen}, \citenamefont {Steinmann}, \citenamefont
  {Peng},\ and\ \citenamefont {Yang}}]{pprpa2}%
  \BibitemOpen
  \bibfield  {author} {\bibinfo {author} {\bibfnamefont {Y.}~\bibnamefont
  {Yang}}, \bibinfo {author} {\bibfnamefont {H.}~\bibnamefont {van Aggelen}},
  \bibinfo {author} {\bibfnamefont {S.~N.}\ \bibnamefont {Steinmann}}, \bibinfo
  {author} {\bibfnamefont {D.}~\bibnamefont {Peng}}, \ and\ \bibinfo {author}
  {\bibfnamefont {W.}~\bibnamefont {Yang}},\ }\href@noop {} {\bibfield
  {journal} {\bibinfo  {journal} {J. Chem. Phys.}\ }\textbf {\bibinfo {volume}
  {139}},\ \bibinfo {pages} {174110} (\bibinfo {year} {2013})}\BibitemShut
  {NoStop}%
\bibitem [{\citenamefont {Zhang}\ \emph {et~al.}(2015)\citenamefont {Zhang},
  \citenamefont {Peng}, \citenamefont {Zhang},\ and\ \citenamefont
  {Yang}}]{pprpa3}%
  \BibitemOpen
  \bibfield  {author} {\bibinfo {author} {\bibfnamefont {D.}~\bibnamefont
  {Zhang}}, \bibinfo {author} {\bibfnamefont {D.}~\bibnamefont {Peng}},
  \bibinfo {author} {\bibfnamefont {P.}~\bibnamefont {Zhang}}, \ and\ \bibinfo
  {author} {\bibfnamefont {W.}~\bibnamefont {Yang}},\ }\href@noop {} {\bibfield
   {journal} {\bibinfo  {journal} {Phys. Chem. Chem. Phys.}\ }\textbf {\bibinfo
  {volume} {17}},\ \bibinfo {pages} {1025} (\bibinfo {year}
  {2015})}\BibitemShut {NoStop}%
\bibitem [{\citenamefont {Peng}\ \emph {et~al.}(2012)\citenamefont {Peng},
  \citenamefont {Hu}, \citenamefont {Devarajan}, \citenamefont {Ess},
  \citenamefont {Johnson},\ and\ \citenamefont {Yang}}]{VFS}%
  \BibitemOpen
  \bibfield  {author} {\bibinfo {author} {\bibfnamefont {D.}~\bibnamefont
  {Peng}}, \bibinfo {author} {\bibfnamefont {X.}~\bibnamefont {Hu}}, \bibinfo
  {author} {\bibfnamefont {D.}~\bibnamefont {Devarajan}}, \bibinfo {author}
  {\bibfnamefont {D.~H.}\ \bibnamefont {Ess}}, \bibinfo {author} {\bibfnamefont
  {E.~R.}\ \bibnamefont {Johnson}}, \ and\ \bibinfo {author} {\bibfnamefont
  {W.}~\bibnamefont {Yang}},\ }\href@noop {} {\bibfield  {journal} {\bibinfo
  {journal} {J. Chem. Phys.}\ }\textbf {\bibinfo {volume} {137}},\ \bibinfo
  {pages} {114112} (\bibinfo {year} {2012})}\BibitemShut {NoStop}%
\bibitem [{\citenamefont {Ess}\ \emph {et~al.}(2011)\citenamefont {Ess},
  \citenamefont {Johnson}, \citenamefont {Hu},\ and\ \citenamefont
  {Yang}}]{FS1}%
  \BibitemOpen
  \bibfield  {author} {\bibinfo {author} {\bibfnamefont {D.~H.}\ \bibnamefont
  {Ess}}, \bibinfo {author} {\bibfnamefont {E.~R.}\ \bibnamefont {Johnson}},
  \bibinfo {author} {\bibfnamefont {X.}~\bibnamefont {Hu}}, \ and\ \bibinfo
  {author} {\bibfnamefont {W.}~\bibnamefont {Yang}},\ }\href@noop {} {\bibfield
   {journal} {\bibinfo  {journal} {J. Phys. Chem. A}\ }\textbf {\bibinfo
  {volume} {115}},\ \bibinfo {pages} {76} (\bibinfo {year} {2011})}\BibitemShut
  {NoStop}%
\bibitem [{\citenamefont {Su}\ \emph {et~al.}(2018)\citenamefont {Su},
  \citenamefont {Li},\ and\ \citenamefont {Yang}}]{FS2}%
  \BibitemOpen
  \bibfield  {author} {\bibinfo {author} {\bibfnamefont {N.~Q.}\ \bibnamefont
  {Su}}, \bibinfo {author} {\bibfnamefont {C.}~\bibnamefont {Li}}, \ and\
  \bibinfo {author} {\bibfnamefont {W.}~\bibnamefont {Yang}},\ }\href@noop {}
  {\bibfield  {journal} {\bibinfo  {journal} {Proc. Natl. Acad. Sci. U.S.A}\
  }\textbf {\bibinfo {volume} {115}},\ \bibinfo {pages} {9678} (\bibinfo {year}
  {2018})}\BibitemShut {NoStop}%
\bibitem [{\citenamefont {Bao}\ \emph {et~al.}(2016)\citenamefont {Bao},
  \citenamefont {Sand}, \citenamefont {Gagliardi},\ and\ \citenamefont
  {Truhlar}}]{PDFT-MG}%
  \BibitemOpen
  \bibfield  {author} {\bibinfo {author} {\bibfnamefont {J.~L.}\ \bibnamefont
  {Bao}}, \bibinfo {author} {\bibfnamefont {A.}~\bibnamefont {Sand}}, \bibinfo
  {author} {\bibfnamefont {L.}~\bibnamefont {Gagliardi}}, \ and\ \bibinfo
  {author} {\bibfnamefont {D.~G.}\ \bibnamefont {Truhlar}},\ }\href@noop {}
  {\bibfield  {journal} {\bibinfo  {journal} {J. Chem. Theory Comput}\ }\textbf
  {\bibinfo {volume} {12}},\ \bibinfo {pages} {4274} (\bibinfo {year}
  {2016})}\BibitemShut {NoStop}%
\bibitem [{\citenamefont {Li~Manni}\ \emph {et~al.}(2014)\citenamefont
  {Li~Manni}, \citenamefont {Carlson}, \citenamefont {Luo}, \citenamefont {Ma},
  \citenamefont {Olsen}, \citenamefont {Truhlar},\ and\ \citenamefont
  {Gagliardi}}]{PDFT1}%
  \BibitemOpen
  \bibfield  {author} {\bibinfo {author} {\bibfnamefont {G.}~\bibnamefont
  {Li~Manni}}, \bibinfo {author} {\bibfnamefont {R.~K.}\ \bibnamefont
  {Carlson}}, \bibinfo {author} {\bibfnamefont {S.}~\bibnamefont {Luo}},
  \bibinfo {author} {\bibfnamefont {D.}~\bibnamefont {Ma}}, \bibinfo {author}
  {\bibfnamefont {J.}~\bibnamefont {Olsen}}, \bibinfo {author} {\bibfnamefont
  {D.~G.}\ \bibnamefont {Truhlar}}, \ and\ \bibinfo {author} {\bibfnamefont
  {L.}~\bibnamefont {Gagliardi}},\ }\href@noop {} {\bibfield  {journal}
  {\bibinfo  {journal} {J. Chem. Theory Comput.}\ }\textbf {\bibinfo {volume}
  {10}},\ \bibinfo {pages} {3669} (\bibinfo {year} {2014})}\BibitemShut
  {NoStop}%
\bibitem [{\citenamefont {Gagliardi}\ \emph {et~al.}(2017)\citenamefont
  {Gagliardi}, \citenamefont {Truhlar}, \citenamefont {Li~Manni}, \citenamefont
  {Carlson}, \citenamefont {Hoyer},\ and\ \citenamefont {Bao}}]{PDFT2}%
  \BibitemOpen
  \bibfield  {author} {\bibinfo {author} {\bibfnamefont {L.}~\bibnamefont
  {Gagliardi}}, \bibinfo {author} {\bibfnamefont {D.~G.}\ \bibnamefont
  {Truhlar}}, \bibinfo {author} {\bibfnamefont {G.}~\bibnamefont {Li~Manni}},
  \bibinfo {author} {\bibfnamefont {R.~K.}\ \bibnamefont {Carlson}}, \bibinfo
  {author} {\bibfnamefont {C.~E.}\ \bibnamefont {Hoyer}}, \ and\ \bibinfo
  {author} {\bibfnamefont {J.~L.}\ \bibnamefont {Bao}},\ }\href@noop {}
  {\bibfield  {journal} {\bibinfo  {journal} {Acc. Chem. Res.}\ }\textbf
  {\bibinfo {volume} {50}},\ \bibinfo {pages} {66} (\bibinfo {year}
  {2017})}\BibitemShut {NoStop}%
\bibitem [{\citenamefont {Casanova}\ and\ \citenamefont
  {Krylov}(2020)}]{SFnew}%
  \BibitemOpen
  \bibfield  {author} {\bibinfo {author} {\bibfnamefont {D.}~\bibnamefont
  {Casanova}}\ and\ \bibinfo {author} {\bibfnamefont {A.~I.}\ \bibnamefont
  {Krylov}},\ }\href@noop {} {\bibfield  {journal} {\bibinfo  {journal} {Phys.
  Chem. Chem. Phys.}\ }\textbf {\bibinfo {volume} {22}},\ \bibinfo {pages}
  {4326} (\bibinfo {year} {2020})}\BibitemShut {NoStop}%
\bibitem [{\citenamefont {Shao}\ \emph {et~al.}(2003)\citenamefont {Shao},
  \citenamefont {Head-Gordon},\ and\ \citenamefont {Krylov}}]{UB3LYP}%
  \BibitemOpen
  \bibfield  {author} {\bibinfo {author} {\bibfnamefont {Y.}~\bibnamefont
  {Shao}}, \bibinfo {author} {\bibfnamefont {M.}~\bibnamefont {Head-Gordon}}, \
  and\ \bibinfo {author} {\bibfnamefont {A.~I.}\ \bibnamefont {Krylov}},\
  }\href@noop {} {\bibfield  {journal} {\bibinfo  {journal} {J. Chem. Phys.}\
  }\textbf {\bibinfo {volume} {118}},\ \bibinfo {pages} {4807} (\bibinfo {year}
  {2003})}\BibitemShut {NoStop}%
\bibitem [{\citenamefont {Bernard}\ \emph {et~al.}(2012)\citenamefont
  {Bernard}, \citenamefont {Shao},\ and\ \citenamefont {Krylov}}]{KrylovSFDFT}%
  \BibitemOpen
  \bibfield  {author} {\bibinfo {author} {\bibfnamefont {Y.~A.}\ \bibnamefont
  {Bernard}}, \bibinfo {author} {\bibfnamefont {Y.}~\bibnamefont {Shao}}, \
  and\ \bibinfo {author} {\bibfnamefont {A.~I.}\ \bibnamefont {Krylov}},\
  }\href@noop {} {\bibfield  {journal} {\bibinfo  {journal} {J. Chem. Phys.}\
  }\textbf {\bibinfo {volume} {136}},\ \bibinfo {pages} {204103} (\bibinfo
  {year} {2012})}\BibitemShut {NoStop}%
\bibitem [{\citenamefont {Pulay}(2011)}]{CASPT2}%
  \BibitemOpen
  \bibfield  {author} {\bibinfo {author} {\bibfnamefont {P.}~\bibnamefont
  {Pulay}},\ }\href@noop {} {\bibfield  {journal} {\bibinfo  {journal} {Int. J.
  Quantum Chem.}\ }\textbf {\bibinfo {volume} {111}},\ \bibinfo {pages} {3273}
  (\bibinfo {year} {2011})}\BibitemShut {NoStop}%
\bibitem [{\citenamefont {Manohar}\ and\ \citenamefont
  {Krylov}(2008)}]{Krylov-SF-CCSDT}%
  \BibitemOpen
  \bibfield  {author} {\bibinfo {author} {\bibfnamefont {P.~U.}\ \bibnamefont
  {Manohar}}\ and\ \bibinfo {author} {\bibfnamefont {A.~I.}\ \bibnamefont
  {Krylov}},\ }\href@noop {} {\bibfield  {journal} {\bibinfo  {journal} {J.
  Chem. Phys.}\ }\textbf {\bibinfo {volume} {129}},\ \bibinfo {pages} {194105}
  (\bibinfo {year} {2008})}\BibitemShut {NoStop}%
\bibitem [{\citenamefont {Slipchenko}\ and\ \citenamefont
  {Krylov}(2002)}]{Krylov-SFCCSD}%
  \BibitemOpen
  \bibfield  {author} {\bibinfo {author} {\bibfnamefont {L.~V.}\ \bibnamefont
  {Slipchenko}}\ and\ \bibinfo {author} {\bibfnamefont {A.~I.}\ \bibnamefont
  {Krylov}},\ }\href@noop {} {\bibfield  {journal} {\bibinfo  {journal} {J.
  Chem. Phys.}\ }\textbf {\bibinfo {volume} {117}},\ \bibinfo {pages} {4694}
  (\bibinfo {year} {2002})}\BibitemShut {NoStop}%
\bibitem [{\citenamefont {Krylov}(2001)}]{SFCI}%
  \BibitemOpen
  \bibfield  {author} {\bibinfo {author} {\bibfnamefont {A.~I.}\ \bibnamefont
  {Krylov}},\ }\href@noop {} {\bibfield  {journal} {\bibinfo  {journal} {Chem.
  Phys. Lett.}\ }\textbf {\bibinfo {volume} {350}},\ \bibinfo {pages} {522 }
  (\bibinfo {year} {2001})}\BibitemShut {NoStop}%
\bibitem [{\citenamefont {Lopez}\ \emph {et~al.}(2011)\citenamefont {Lopez},
  \citenamefont {Ruipérez}, \citenamefont {Piris}, \citenamefont {Matxain},\
  and\ \citenamefont {Ugalde}}]{NOFT-Birad}%
  \BibitemOpen
  \bibfield  {author} {\bibinfo {author} {\bibfnamefont {X.}~\bibnamefont
  {Lopez}}, \bibinfo {author} {\bibfnamefont {F.}~\bibnamefont {Ruipérez}},
  \bibinfo {author} {\bibfnamefont {M.}~\bibnamefont {Piris}}, \bibinfo
  {author} {\bibfnamefont {J.~M.}\ \bibnamefont {Matxain}}, \ and\ \bibinfo
  {author} {\bibfnamefont {J.~M.}\ \bibnamefont {Ugalde}},\ }\href@noop {}
  {\bibfield  {journal} {\bibinfo  {journal} {ChemPhysChem}\ }\textbf {\bibinfo
  {volume} {12}},\ \bibinfo {pages} {1061} (\bibinfo {year}
  {2011})}\BibitemShut {NoStop}%
\bibitem [{\citenamefont {Shee}\ \emph {et~al.}(2019)\citenamefont {Shee},
  \citenamefont {Arthur}, \citenamefont {Zhang}, \citenamefont {Reichman},\
  and\ \citenamefont {Friesner}}]{RF-AFQMC}%
  \BibitemOpen
  \bibfield  {author} {\bibinfo {author} {\bibfnamefont {J.}~\bibnamefont
  {Shee}}, \bibinfo {author} {\bibfnamefont {E.~J.}\ \bibnamefont {Arthur}},
  \bibinfo {author} {\bibfnamefont {S.}~\bibnamefont {Zhang}}, \bibinfo
  {author} {\bibfnamefont {D.~R.}\ \bibnamefont {Reichman}}, \ and\ \bibinfo
  {author} {\bibfnamefont {R.~A.}\ \bibnamefont {Friesner}},\ }\href@noop {}
  {\bibfield  {journal} {\bibinfo  {journal} {J. Chem. Theory Comput.}\
  }\textbf {\bibinfo {volume} {15}},\ \bibinfo {pages} {4924} (\bibinfo {year}
  {2019})}\BibitemShut {NoStop}%
\bibitem [{\citenamefont {Piris}(2019)}]{PirisSpinAvg}%
  \BibitemOpen
  \bibfield  {author} {\bibinfo {author} {\bibfnamefont {M.}~\bibnamefont
  {Piris}},\ }\href {\doibase 10.1103/PhysRevA.100.032508} {\bibfield
  {journal} {\bibinfo  {journal} {Phys. Rev. A}\ }\textbf {\bibinfo {volume}
  {100}},\ \bibinfo {pages} {032508} (\bibinfo {year} {2019})}\BibitemShut
  {NoStop}%
\bibitem [{\citenamefont {Gidofalvi}\ and\ \citenamefont
  {Mazziotti}(2005)}]{Gidofalvi2005}%
  \BibitemOpen
  \bibfield  {author} {\bibinfo {author} {\bibfnamefont {G.}~\bibnamefont
  {Gidofalvi}}\ and\ \bibinfo {author} {\bibfnamefont {D.~A.}\ \bibnamefont
  {Mazziotti}},\ }\href {\doibase 10.1103/physreva.72.052505} {\bibfield
  {journal} {\bibinfo  {journal} {Phys. Rev. A}\ }\textbf {\bibinfo {volume}
  {72}} (\bibinfo {year} {2005}),\ 10.1103/physreva.72.052505}\BibitemShut
  {NoStop}%
\bibitem [{\citenamefont {P{\'e}rez-Romero}\ \emph {et~al.}(1997)\citenamefont
  {P{\'e}rez-Romero}, \citenamefont {Tel},\ and\ \citenamefont
  {Valdemoro}}]{Perez-Romero1997}%
  \BibitemOpen
  \bibfield  {author} {\bibinfo {author} {\bibfnamefont {E.}~\bibnamefont
  {P{\'e}rez-Romero}}, \bibinfo {author} {\bibfnamefont {L.~M.}\ \bibnamefont
  {Tel}}, \ and\ \bibinfo {author} {\bibfnamefont {C.}~\bibnamefont
  {Valdemoro}},\ }\href {\doibase
  10.1002/(sici)1097-461x(1997)61:1<55::aid-qua6>3.0.co;2-3} {\bibfield
  {journal} {\bibinfo  {journal} {Int. J. Quantum Chem.}\ }\textbf {\bibinfo
  {volume} {61}},\ \bibinfo {pages} {55} (\bibinfo {year} {1997})}\BibitemShut
  {NoStop}%
\bibitem [{\citenamefont {Mihailovi{\'{c}}}\ and\ \citenamefont
  {Rosina}(1975)}]{Mihailovic1975}%
  \BibitemOpen
  \bibfield  {author} {\bibinfo {author} {\bibfnamefont {M.}~\bibnamefont
  {Mihailovi{\'{c}}}}\ and\ \bibinfo {author} {\bibfnamefont {M.}~\bibnamefont
  {Rosina}},\ }\href {\doibase 10.1016/0375-9474(75)90420-0} {\bibfield
  {journal} {\bibinfo  {journal} {Nucl. Phys. A}\ }\textbf {\bibinfo {volume}
  {237}},\ \bibinfo {pages} {221} (\bibinfo {year} {1975})}\BibitemShut
  {NoStop}%
\bibitem [{\citenamefont {Mazziotti}(2006)}]{ACSE1}%
  \BibitemOpen
  \bibfield  {author} {\bibinfo {author} {\bibfnamefont {D.~A.}\ \bibnamefont
  {Mazziotti}},\ }\href@noop {} {\bibfield  {journal} {\bibinfo  {journal}
  {Phys. Rev. Lett.}\ }\textbf {\bibinfo {volume} {97}},\ \bibinfo {pages}
  {143002} (\bibinfo {year} {2006})}\BibitemShut {NoStop}%
\bibitem [{\citenamefont {Mazziotti}(2007{\natexlab{a}})}]{ACSE2}%
  \BibitemOpen
  \bibfield  {author} {\bibinfo {author} {\bibfnamefont {D.~A.}\ \bibnamefont
  {Mazziotti}},\ }\href@noop {} {\bibfield  {journal} {\bibinfo  {journal}
  {Phys. Rev. A}\ }\textbf {\bibinfo {volume} {75}},\ \bibinfo {pages} {022505}
  (\bibinfo {year} {2007}{\natexlab{a}})}\BibitemShut {NoStop}%
\bibitem [{\citenamefont {Mazziotti}(2007{\natexlab{b}})}]{ACSEcor}%
  \BibitemOpen
  \bibfield  {author} {\bibinfo {author} {\bibfnamefont {D.~A.}\ \bibnamefont
  {Mazziotti}},\ }\href@noop {} {\bibfield  {journal} {\bibinfo  {journal}
  {Phys. Rev. A}\ }\textbf {\bibinfo {volume} {76}},\ \bibinfo {pages} {052502}
  (\bibinfo {year} {2007}{\natexlab{b}})}\BibitemShut {NoStop}%
\bibitem [{\citenamefont {Mazziotti}(2008)}]{ACSEexp1}%
  \BibitemOpen
  \bibfield  {author} {\bibinfo {author} {\bibfnamefont {D.~A.}\ \bibnamefont
  {Mazziotti}},\ }\href@noop {} {\bibfield  {journal} {\bibinfo  {journal} {J.
  Phys. Chem. A}\ }\textbf {\bibinfo {volume} {112}},\ \bibinfo {pages} {13684}
  (\bibinfo {year} {2008})}\BibitemShut {NoStop}%
\bibitem [{\citenamefont {Sand}\ and\ \citenamefont
  {Mazziotti}(2015{\natexlab{a}})}]{ACSEexp2}%
  \BibitemOpen
  \bibfield  {author} {\bibinfo {author} {\bibfnamefont {A.~M.}\ \bibnamefont
  {Sand}}\ and\ \bibinfo {author} {\bibfnamefont {D.~A.}\ \bibnamefont
  {Mazziotti}},\ }\href@noop {} {\bibfield  {journal} {\bibinfo  {journal} {J.
  Chem. Phys.}\ }\textbf {\bibinfo {volume} {143}},\ \bibinfo {pages} {134110}
  (\bibinfo {year} {2015}{\natexlab{a}})}\BibitemShut {NoStop}%
\bibitem [{\citenamefont {Mazziotti}(1998{\natexlab{a}})}]{CSE}%
  \BibitemOpen
  \bibfield  {author} {\bibinfo {author} {\bibfnamefont {D.~A.}\ \bibnamefont
  {Mazziotti}},\ }\href@noop {} {\bibfield  {journal} {\bibinfo  {journal}
  {Phys. Rev. A}\ }\textbf {\bibinfo {volume} {57}},\ \bibinfo {pages} {4219}
  (\bibinfo {year} {1998}{\natexlab{a}})}\BibitemShut {NoStop}%
\bibitem [{\citenamefont {DePrince}\ and\ \citenamefont
  {Mazziotti}(2007)}]{cumulant}%
  \BibitemOpen
  \bibfield  {author} {\bibinfo {author} {\bibfnamefont {A.~E.}\ \bibnamefont
  {DePrince}}\ and\ \bibinfo {author} {\bibfnamefont {D.~A.}\ \bibnamefont
  {Mazziotti}},\ }\href@noop {} {\bibfield  {journal} {\bibinfo  {journal} {J.
  Chem. Phys.}\ }\textbf {\bibinfo {volume} {127}},\ \bibinfo {pages} {104104}
  (\bibinfo {year} {2007})}\BibitemShut {NoStop}%
\bibitem [{\citenamefont {Mazziotti}(1998{\natexlab{b}})}]{schwinger}%
  \BibitemOpen
  \bibfield  {author} {\bibinfo {author} {\bibfnamefont {D.~A.}\ \bibnamefont
  {Mazziotti}},\ }\href@noop {} {\bibfield  {journal} {\bibinfo  {journal}
  {Chem. Phys. Lett.}\ }\textbf {\bibinfo {volume} {289}},\ \bibinfo {pages}
  {419 } (\bibinfo {year} {1998}{\natexlab{b}})}\BibitemShut {NoStop}%
\bibitem [{\citenamefont {Mazziotti}(1998{\natexlab{c}})}]{CSE35}%
  \BibitemOpen
  \bibfield  {author} {\bibinfo {author} {\bibfnamefont {D.~A.}\ \bibnamefont
  {Mazziotti}},\ }\href@noop {} {\bibfield  {journal} {\bibinfo  {journal}
  {Int. J. Quantum Chem.}\ }\textbf {\bibinfo {volume} {70}},\ \bibinfo {pages}
  {557} (\bibinfo {year} {1998}{\natexlab{c}})}\BibitemShut {NoStop}%
\bibitem [{\citenamefont {Rothman}\ \emph {et~al.}(2009)\citenamefont
  {Rothman}, \citenamefont {Foley},\ and\ \citenamefont
  {Mazziotti}}]{ACSEspin}%
  \BibitemOpen
  \bibfield  {author} {\bibinfo {author} {\bibfnamefont {A.~E.}\ \bibnamefont
  {Rothman}}, \bibinfo {author} {\bibfnamefont {J.~J.}\ \bibnamefont {Foley}},
  \ and\ \bibinfo {author} {\bibfnamefont {D.~A.}\ \bibnamefont {Mazziotti}},\
  }\href@noop {} {\bibfield  {journal} {\bibinfo  {journal} {Phys. Rev. A}\
  }\textbf {\bibinfo {volume} {80}},\ \bibinfo {pages} {052508} (\bibinfo
  {year} {2009})}\BibitemShut {NoStop}%
\bibitem [{\citenamefont {{\relax Maplesoft, a division of Waterloo Maple Inc.,
  Waterloo, Ontario.}}(2019)}]{Maple}%
  \BibitemOpen
  \bibfield  {author} {\bibinfo {author} {\bibnamefont {{\relax Maplesoft, a
  division of Waterloo Maple Inc., Waterloo, Ontario.}}},\ }\href@noop {} {\
  (\bibinfo {year} {2019})}\BibitemShut {NoStop}%
\bibitem [{\citenamefont {{RDMChem, Chicago, Illinois.}}(2019)}]{QCP}%
  \BibitemOpen
  \bibfield  {author} {\bibinfo {author} {\bibnamefont {{RDMChem, Chicago,
  Illinois.}}},\ }\href@noop {} {\  (\bibinfo {year} {2019})}\BibitemShut
  {NoStop}%
\bibitem [{\citenamefont {Sand}\ and\ \citenamefont
  {Mazziotti}(2015{\natexlab{b}})}]{ACSEextrap}%
  \BibitemOpen
  \bibfield  {author} {\bibinfo {author} {\bibfnamefont {A.~M.}\ \bibnamefont
  {Sand}}\ and\ \bibinfo {author} {\bibfnamefont {D.~A.}\ \bibnamefont
  {Mazziotti}},\ }\href@noop {} {\bibfield  {journal} {\bibinfo  {journal} {J.
  Chem. Phys.}\ }\textbf {\bibinfo {volume} {143}},\ \bibinfo {pages} {134110}
  (\bibinfo {year} {2015}{\natexlab{b}})}\BibitemShut {NoStop}%
\bibitem [{\citenamefont {Herzberg}(1979)}]{HuberConst}%
  \BibitemOpen
  \bibfield  {author} {\bibinfo {author} {\bibfnamefont {K.}~\bibnamefont
  {Herzberg}, \bibfnamefont {G.;~Huber}},\ }\href@noop {} {\bibfield  {journal}
  {\bibinfo  {journal} {Springer Verlag}\ } (\bibinfo {year}
  {1979})}\BibitemShut {NoStop}%
\bibitem [{\citenamefont {Osmann}\ \emph {et~al.}(1997)\citenamefont {Osmann},
  \citenamefont {Bunker}, \citenamefont {Jensen},\ and\ \citenamefont
  {Kraemer}}]{NH2}%
  \BibitemOpen
  \bibfield  {author} {\bibinfo {author} {\bibfnamefont {G.}~\bibnamefont
  {Osmann}}, \bibinfo {author} {\bibfnamefont {P.}~\bibnamefont {Bunker}},
  \bibinfo {author} {\bibfnamefont {P.}~\bibnamefont {Jensen}}, \ and\ \bibinfo
  {author} {\bibfnamefont {W.}~\bibnamefont {Kraemer}},\ }\href@noop {}
  {\bibfield  {journal} {\bibinfo  {journal} {J. Mol. Spectrosc.}\ }\textbf
  {\bibinfo {volume} {186}},\ \bibinfo {pages} {319 } (\bibinfo {year}
  {1997})}\BibitemShut {NoStop}%
\bibitem [{\citenamefont {Gu}\ \emph {et~al.}(2000)\citenamefont {Gu},
  \citenamefont {Hirsch}, \citenamefont {Buenker}, \citenamefont {Brumm},
  \citenamefont {Osmann}, \citenamefont {Bunker},\ and\ \citenamefont
  {Jensen}}]{CH2}%
  \BibitemOpen
  \bibfield  {author} {\bibinfo {author} {\bibfnamefont {J.-P.}\ \bibnamefont
  {Gu}}, \bibinfo {author} {\bibfnamefont {G.}~\bibnamefont {Hirsch}}, \bibinfo
  {author} {\bibfnamefont {R.}~\bibnamefont {Buenker}}, \bibinfo {author}
  {\bibfnamefont {M.}~\bibnamefont {Brumm}}, \bibinfo {author} {\bibfnamefont
  {G.}~\bibnamefont {Osmann}}, \bibinfo {author} {\bibfnamefont
  {P.}~\bibnamefont {Bunker}}, \ and\ \bibinfo {author} {\bibfnamefont
  {P.}~\bibnamefont {Jensen}},\ }\href@noop {} {\bibfield  {journal} {\bibinfo
  {journal} {J. Mol. Struct.}\ }\textbf {\bibinfo {volume} {517-518}},\
  \bibinfo {pages} {247 } (\bibinfo {year} {2000})}\BibitemShut {NoStop}%
\bibitem [{\citenamefont {Berkowitz}\ and\ \citenamefont {Cho}(1989)}]{PH2}%
  \BibitemOpen
  \bibfield  {author} {\bibinfo {author} {\bibfnamefont {J.}~\bibnamefont
  {Berkowitz}}\ and\ \bibinfo {author} {\bibfnamefont {H.}~\bibnamefont
  {Cho}},\ }\href@noop {} {\bibfield  {journal} {\bibinfo  {journal} {J. Chem.
  Phys.}\ }\textbf {\bibinfo {volume} {90}},\ \bibinfo {pages} {1} (\bibinfo
  {year} {1989})}\BibitemShut {NoStop}%
\bibitem [{\citenamefont {Berkowitz}\ \emph {et~al.}(1987)\citenamefont
  {Berkowitz}, \citenamefont {Greene}, \citenamefont {Cho},\ and\ \citenamefont
  {Ruščić}}]{SiH2}%
  \BibitemOpen
  \bibfield  {author} {\bibinfo {author} {\bibfnamefont {J.}~\bibnamefont
  {Berkowitz}}, \bibinfo {author} {\bibfnamefont {J.~P.}\ \bibnamefont
  {Greene}}, \bibinfo {author} {\bibfnamefont {H.}~\bibnamefont {Cho}}, \ and\
  \bibinfo {author} {\bibfnamefont {B.}~\bibnamefont {Ruščić}},\ }\href@noop
  {} {\bibfield  {journal} {\bibinfo  {journal} {J. Chem. Phys.}\ }\textbf
  {\bibinfo {volume} {86}},\ \bibinfo {pages} {1235} (\bibinfo {year}
  {1987})}\BibitemShut {NoStop}%
\bibitem [{\citenamefont {Dunning}(1989)}]{ccbasis}%
  \BibitemOpen
  \bibfield  {author} {\bibinfo {author} {\bibfnamefont {T.~H.}\ \bibnamefont
  {Dunning}},\ }\href@noop {} {\bibfield  {journal} {\bibinfo  {journal} {J.
  Chem. Phys.}\ }\textbf {\bibinfo {volume} {90}},\ \bibinfo {pages} {1007}
  (\bibinfo {year} {1989})}\BibitemShut {NoStop}%
\bibitem [{\citenamefont {Woon}\ and\ \citenamefont
  {Dunning}(1993)}]{ccbasis2}%
  \BibitemOpen
  \bibfield  {author} {\bibinfo {author} {\bibfnamefont {D.~E.}\ \bibnamefont
  {Woon}}\ and\ \bibinfo {author} {\bibfnamefont {T.~H.}\ \bibnamefont
  {Dunning}},\ }\href@noop {} {\bibfield  {journal} {\bibinfo  {journal} {J.
  Chem. Phys.}\ }\textbf {\bibinfo {volume} {98}},\ \bibinfo {pages} {1358}
  (\bibinfo {year} {1993})}\BibitemShut {NoStop}%
\bibitem [{\citenamefont {Kawamura}\ \emph {et~al.}(2020)\citenamefont
  {Kawamura}, \citenamefont {Xie}, \citenamefont {Boyn}, \citenamefont {Jesse},
  \citenamefont {McNeece}, \citenamefont {Hill}, \citenamefont {Collins},
  \citenamefont {Valdez-Moreira}, \citenamefont {Filatov}, \citenamefont
  {Kurutz}, \citenamefont {Mazziotti},\ and\ \citenamefont
  {Anderson}}]{FeTTfttdirad}%
  \BibitemOpen
  \bibfield  {author} {\bibinfo {author} {\bibfnamefont {A.}~\bibnamefont
  {Kawamura}}, \bibinfo {author} {\bibfnamefont {J.}~\bibnamefont {Xie}},
  \bibinfo {author} {\bibfnamefont {J.-N.}\ \bibnamefont {Boyn}}, \bibinfo
  {author} {\bibfnamefont {K.~A.}\ \bibnamefont {Jesse}}, \bibinfo {author}
  {\bibfnamefont {A.~J.}\ \bibnamefont {McNeece}}, \bibinfo {author}
  {\bibfnamefont {E.~A.}\ \bibnamefont {Hill}}, \bibinfo {author}
  {\bibfnamefont {K.~A.}\ \bibnamefont {Collins}}, \bibinfo {author}
  {\bibfnamefont {J.~A.}\ \bibnamefont {Valdez-Moreira}}, \bibinfo {author}
  {\bibfnamefont {A.~S.}\ \bibnamefont {Filatov}}, \bibinfo {author}
  {\bibfnamefont {J.~W.}\ \bibnamefont {Kurutz}}, \bibinfo {author}
  {\bibfnamefont {D.~A.}\ \bibnamefont {Mazziotti}}, \ and\ \bibinfo {author}
  {\bibfnamefont {J.~S.}\ \bibnamefont {Anderson}},\ }\href@noop {} {\bibfield
  {journal} {\bibinfo  {journal} {J. Am. Chem. Soc.}\ }\textbf {\bibinfo
  {volume} {142}},\ \bibinfo {pages} {17670} (\bibinfo {year}
  {2020})}\BibitemShut {NoStop}%
\bibitem [{\citenamefont {Gallagher}\ \emph {et~al.}(2016)\citenamefont
  {Gallagher}, \citenamefont {Bauer}, \citenamefont {Pink}, \citenamefont
  {Rajca},\ and\ \citenamefont {Rajca}}]{OrgBirad}%
  \BibitemOpen
  \bibfield  {author} {\bibinfo {author} {\bibfnamefont {N.~M.}\ \bibnamefont
  {Gallagher}}, \bibinfo {author} {\bibfnamefont {J.~J.}\ \bibnamefont
  {Bauer}}, \bibinfo {author} {\bibfnamefont {M.}~\bibnamefont {Pink}},
  \bibinfo {author} {\bibfnamefont {S.}~\bibnamefont {Rajca}}, \ and\ \bibinfo
  {author} {\bibfnamefont {A.}~\bibnamefont {Rajca}},\ }\href@noop {}
  {\bibfield  {journal} {\bibinfo  {journal} {J. Am. Chem. Soc.}\ }\textbf
  {\bibinfo {volume} {138}},\ \bibinfo {pages} {9377} (\bibinfo {year}
  {2016})}\BibitemShut {NoStop}%
\bibitem [{\citenamefont {Yang}\ \emph {et~al.}(2020)\citenamefont {Yang},
  \citenamefont {Zhang}, \citenamefont {Harbuzaru}, \citenamefont {Wang},
  \citenamefont {Wang}, \citenamefont {Koh}, \citenamefont {Guo}, \citenamefont
  {Shi}, \citenamefont {Chen}, \citenamefont {Sun}, \citenamefont {Feng},
  \citenamefont {Ruiz~Delgado}, \citenamefont {Woo}, \citenamefont {Ortiz},\
  and\ \citenamefont {Guo}}]{OrgBirad2}%
  \BibitemOpen
  \bibfield  {author} {\bibinfo {author} {\bibfnamefont {K.}~\bibnamefont
  {Yang}}, \bibinfo {author} {\bibfnamefont {X.}~\bibnamefont {Zhang}},
  \bibinfo {author} {\bibfnamefont {A.}~\bibnamefont {Harbuzaru}}, \bibinfo
  {author} {\bibfnamefont {L.}~\bibnamefont {Wang}}, \bibinfo {author}
  {\bibfnamefont {Y.}~\bibnamefont {Wang}}, \bibinfo {author} {\bibfnamefont
  {C.}~\bibnamefont {Koh}}, \bibinfo {author} {\bibfnamefont {H.}~\bibnamefont
  {Guo}}, \bibinfo {author} {\bibfnamefont {Y.}~\bibnamefont {Shi}}, \bibinfo
  {author} {\bibfnamefont {J.}~\bibnamefont {Chen}}, \bibinfo {author}
  {\bibfnamefont {H.}~\bibnamefont {Sun}}, \bibinfo {author} {\bibfnamefont
  {K.}~\bibnamefont {Feng}}, \bibinfo {author} {\bibfnamefont {M.~C.}\
  \bibnamefont {Ruiz~Delgado}}, \bibinfo {author} {\bibfnamefont {H.~Y.}\
  \bibnamefont {Woo}}, \bibinfo {author} {\bibfnamefont {R.~P.}\ \bibnamefont
  {Ortiz}}, \ and\ \bibinfo {author} {\bibfnamefont {X.}~\bibnamefont {Guo}},\
  }\href@noop {} {\bibfield  {journal} {\bibinfo  {journal} {J. Am. Chem.
  Soc.}\ }\textbf {\bibinfo {volume} {142}},\ \bibinfo {pages} {4329} (\bibinfo
  {year} {2020})}\BibitemShut {NoStop}%
\bibitem [{\citenamefont {Rajca}(1994)}]{OrgBirad3}%
  \BibitemOpen
  \bibfield  {author} {\bibinfo {author} {\bibfnamefont {A.}~\bibnamefont
  {Rajca}},\ }\href@noop {} {\bibfield  {journal} {\bibinfo  {journal} {Chem.
  Rev.}\ }\textbf {\bibinfo {volume} {94}},\ \bibinfo {pages} {871} (\bibinfo
  {year} {1994})}\BibitemShut {NoStop}%
\bibitem [{\citenamefont {Rudebusch}\ \emph {et~al.}(2016)\citenamefont
  {Rudebusch}, \citenamefont {Zafra}, \citenamefont {Jorner}, \citenamefont
  {Fukuda}, \citenamefont {Marshall}, \citenamefont {Arrechea-Marcos},
  \citenamefont {Espejo}, \citenamefont {Ponce~Ortiz}, \citenamefont
  {G{\'o}mez-Garc{\'i}a}, \citenamefont {Zakharov}, \citenamefont {Nakano},
  \citenamefont {Ottosson}, \citenamefont {Casado},\ and\ \citenamefont
  {Haley}}]{OrgBirad4}%
  \BibitemOpen
  \bibfield  {author} {\bibinfo {author} {\bibfnamefont {G.~E.}\ \bibnamefont
  {Rudebusch}}, \bibinfo {author} {\bibfnamefont {J.~L.}\ \bibnamefont
  {Zafra}}, \bibinfo {author} {\bibfnamefont {K.}~\bibnamefont {Jorner}},
  \bibinfo {author} {\bibfnamefont {K.}~\bibnamefont {Fukuda}}, \bibinfo
  {author} {\bibfnamefont {J.~L.}\ \bibnamefont {Marshall}}, \bibinfo {author}
  {\bibfnamefont {I.}~\bibnamefont {Arrechea-Marcos}}, \bibinfo {author}
  {\bibfnamefont {G.~L.}\ \bibnamefont {Espejo}}, \bibinfo {author}
  {\bibfnamefont {R.}~\bibnamefont {Ponce~Ortiz}}, \bibinfo {author}
  {\bibfnamefont {C.~J.}\ \bibnamefont {G{\'o}mez-Garc{\'i}a}}, \bibinfo
  {author} {\bibfnamefont {L.~N.}\ \bibnamefont {Zakharov}}, \bibinfo {author}
  {\bibfnamefont {M.}~\bibnamefont {Nakano}}, \bibinfo {author} {\bibfnamefont
  {H.}~\bibnamefont {Ottosson}}, \bibinfo {author} {\bibfnamefont
  {J.}~\bibnamefont {Casado}}, \ and\ \bibinfo {author} {\bibfnamefont {M.~M.}\
  \bibnamefont {Haley}},\ }\href@noop {} {\bibfield  {journal} {\bibinfo
  {journal} {Nat. Chem.}\ }\textbf {\bibinfo {volume} {8}},\ \bibinfo {pages}
  {753} (\bibinfo {year} {2016})}\BibitemShut {NoStop}%
\bibitem [{\citenamefont {He}\ \emph {et~al.}(2020)\citenamefont {He},
  \citenamefont {Qiu},\ and\ \citenamefont {Li}}]{BenzynesSynth1}%
  \BibitemOpen
  \bibfield  {author} {\bibinfo {author} {\bibfnamefont {J.}~\bibnamefont
  {He}}, \bibinfo {author} {\bibfnamefont {D.}~\bibnamefont {Qiu}}, \ and\
  \bibinfo {author} {\bibfnamefont {Y.}~\bibnamefont {Li}},\ }\href@noop {}
  {\bibfield  {journal} {\bibinfo  {journal} {Acc. Chem. Res.}\ }\textbf
  {\bibinfo {volume} {53}},\ \bibinfo {pages} {508} (\bibinfo {year}
  {2020})}\BibitemShut {NoStop}%
\bibitem [{\citenamefont {Dubrovskiy}\ \emph {et~al.}(2013)\citenamefont
  {Dubrovskiy}, \citenamefont {Markina},\ and\ \citenamefont
  {Larock}}]{BenzyneSynth2}%
  \BibitemOpen
  \bibfield  {author} {\bibinfo {author} {\bibfnamefont {A.~V.}\ \bibnamefont
  {Dubrovskiy}}, \bibinfo {author} {\bibfnamefont {N.~A.}\ \bibnamefont
  {Markina}}, \ and\ \bibinfo {author} {\bibfnamefont {R.~C.}\ \bibnamefont
  {Larock}},\ }\href@noop {} {\bibfield  {journal} {\bibinfo  {journal} {Org.
  Biomol. Chem.}\ }\textbf {\bibinfo {volume} {11}},\ \bibinfo {pages} {191}
  (\bibinfo {year} {2013})}\BibitemShut {NoStop}%
\bibitem [{\citenamefont {Shi}\ \emph {et~al.}(2008)\citenamefont {Shi},
  \citenamefont {Waldo}, \citenamefont {Chen},\ and\ \citenamefont
  {Larock}}]{BenzyneSynth3}%
  \BibitemOpen
  \bibfield  {author} {\bibinfo {author} {\bibfnamefont {F.}~\bibnamefont
  {Shi}}, \bibinfo {author} {\bibfnamefont {J.~P.}\ \bibnamefont {Waldo}},
  \bibinfo {author} {\bibfnamefont {Y.}~\bibnamefont {Chen}}, \ and\ \bibinfo
  {author} {\bibfnamefont {R.~C.}\ \bibnamefont {Larock}},\ }\href@noop {}
  {\bibfield  {journal} {\bibinfo  {journal} {Org. Lett.}\ }\textbf {\bibinfo
  {volume} {10}},\ \bibinfo {pages} {2409} (\bibinfo {year}
  {2008})}\BibitemShut {NoStop}%
\bibitem [{\citenamefont {Sander}(1999)}]{mpBenzyneRev}%
  \BibitemOpen
  \bibfield  {author} {\bibinfo {author} {\bibfnamefont {W.}~\bibnamefont
  {Sander}},\ }\href@noop {} {\bibfield  {journal} {\bibinfo  {journal} {Acc.
  Chem. Res.}\ }\textbf {\bibinfo {volume} {32}},\ \bibinfo {pages} {669}
  (\bibinfo {year} {1999})}\BibitemShut {NoStop}%
\bibitem [{\citenamefont {Wenthold}\ \emph {et~al.}(1998)\citenamefont
  {Wenthold}, \citenamefont {Squires},\ and\ \citenamefont
  {Lineberger}}]{BenzyneExp}%
  \BibitemOpen
  \bibfield  {author} {\bibinfo {author} {\bibfnamefont {P.~G.}\ \bibnamefont
  {Wenthold}}, \bibinfo {author} {\bibfnamefont {R.~R.}\ \bibnamefont
  {Squires}}, \ and\ \bibinfo {author} {\bibfnamefont {W.~C.}\ \bibnamefont
  {Lineberger}},\ }\href@noop {} {\bibfield  {journal} {\bibinfo  {journal} {J.
  Am. Chem. Soc.}\ }\textbf {\bibinfo {volume} {120}},\ \bibinfo {pages} {5279}
  (\bibinfo {year} {1998})}\BibitemShut {NoStop}%
\bibitem [{\citenamefont {Dowd}(1972)}]{TMM}%
  \BibitemOpen
  \bibfield  {author} {\bibinfo {author} {\bibfnamefont {P.}~\bibnamefont
  {Dowd}},\ }\href@noop {} {\bibfield  {journal} {\bibinfo  {journal} {Acc.
  Chem. Res.}\ }\textbf {\bibinfo {volume} {5}},\ \bibinfo {pages} {242}
  (\bibinfo {year} {1972})}\BibitemShut {NoStop}%
\bibitem [{\citenamefont {Stoneburner}\ \emph {et~al.}(2017)\citenamefont
  {Stoneburner}, \citenamefont {Shen}, \citenamefont {Ajala}, \citenamefont
  {Piecuch}, \citenamefont {Truhlar},\ and\ \citenamefont
  {Gagliardi}}]{Laura_MRCC}%
  \BibitemOpen
  \bibfield  {author} {\bibinfo {author} {\bibfnamefont {S.~J.}\ \bibnamefont
  {Stoneburner}}, \bibinfo {author} {\bibfnamefont {J.}~\bibnamefont {Shen}},
  \bibinfo {author} {\bibfnamefont {A.~O.}\ \bibnamefont {Ajala}}, \bibinfo
  {author} {\bibfnamefont {P.}~\bibnamefont {Piecuch}}, \bibinfo {author}
  {\bibfnamefont {D.~G.}\ \bibnamefont {Truhlar}}, \ and\ \bibinfo {author}
  {\bibfnamefont {L.}~\bibnamefont {Gagliardi}},\ }\href@noop {} {\bibfield
  {journal} {\bibinfo  {journal} {J. Chem. Phys.}\ }\textbf {\bibinfo {volume}
  {147}},\ \bibinfo {pages} {164120} (\bibinfo {year} {2017})}\BibitemShut
  {NoStop}%
\bibitem [{\citenamefont {Saito}\ \emph {et~al.}(2011)\citenamefont {Saito},
  \citenamefont {Nishihara}, \citenamefont {Yamanaka}, \citenamefont
  {Kitagawa}, \citenamefont {Kawakami}, \citenamefont {Yamada}, \citenamefont
  {Isobe}, \citenamefont {Okumura},\ and\ \citenamefont {Yamaguchi}}]{mkCCSD}%
  \BibitemOpen
  \bibfield  {author} {\bibinfo {author} {\bibfnamefont {T.}~\bibnamefont
  {Saito}}, \bibinfo {author} {\bibfnamefont {S.}~\bibnamefont {Nishihara}},
  \bibinfo {author} {\bibfnamefont {S.}~\bibnamefont {Yamanaka}}, \bibinfo
  {author} {\bibfnamefont {Y.}~\bibnamefont {Kitagawa}}, \bibinfo {author}
  {\bibfnamefont {T.}~\bibnamefont {Kawakami}}, \bibinfo {author}
  {\bibfnamefont {S.}~\bibnamefont {Yamada}}, \bibinfo {author} {\bibfnamefont
  {H.}~\bibnamefont {Isobe}}, \bibinfo {author} {\bibfnamefont
  {M.}~\bibnamefont {Okumura}}, \ and\ \bibinfo {author} {\bibfnamefont
  {K.}~\bibnamefont {Yamaguchi}},\ }\href@noop {} {\bibfield  {journal}
  {\bibinfo  {journal} {Theor. Chem. Acc.}\ }\textbf {\bibinfo {volume}
  {130}},\ \bibinfo {pages} {749} (\bibinfo {year} {2011})}\BibitemShut
  {NoStop}%
\bibitem [{\citenamefont {Bally}\ and\ \citenamefont
  {Masamune}(1980)}]{cyclobutadiene}%
  \BibitemOpen
  \bibfield  {author} {\bibinfo {author} {\bibfnamefont {T.}~\bibnamefont
  {Bally}}\ and\ \bibinfo {author} {\bibfnamefont {S.}~\bibnamefont
  {Masamune}},\ }\href@noop {} {\bibfield  {journal} {\bibinfo  {journal}
  {Tetrahedron}\ }\textbf {\bibinfo {volume} {36}},\ \bibinfo {pages} {343 }
  (\bibinfo {year} {1980})}\BibitemShut {NoStop}%
\bibitem [{\citenamefont {Anthony}(2008)}]{Acenes1}%
  \BibitemOpen
  \bibfield  {author} {\bibinfo {author} {\bibfnamefont {J.~E.}\ \bibnamefont
  {Anthony}},\ }\href@noop {} {\bibfield  {journal} {\bibinfo  {journal}
  {Angew. Chem.}\ }\textbf {\bibinfo {volume} {47}},\ \bibinfo {pages} {452}
  (\bibinfo {year} {2008})}\BibitemShut {NoStop}%
\bibitem [{\citenamefont {Anthony}(2006)}]{Acenes2}%
  \BibitemOpen
  \bibfield  {author} {\bibinfo {author} {\bibfnamefont {J.~E.}\ \bibnamefont
  {Anthony}},\ }\href@noop {} {\bibfield  {journal} {\bibinfo  {journal} {Chem.
  Rev.}\ }\textbf {\bibinfo {volume} {106}},\ \bibinfo {pages} {5028} (\bibinfo
  {year} {2006})}\BibitemShut {NoStop}%
\bibitem [{\citenamefont {Shen}\ \emph {et~al.}(2018)\citenamefont {Shen},
  \citenamefont {Tatchen}, \citenamefont {Sanchez-Garcia},\ and\ \citenamefont
  {Bettinger}}]{aceneexp}%
  \BibitemOpen
  \bibfield  {author} {\bibinfo {author} {\bibfnamefont {B.}~\bibnamefont
  {Shen}}, \bibinfo {author} {\bibfnamefont {J.}~\bibnamefont {Tatchen}},
  \bibinfo {author} {\bibfnamefont {E.}~\bibnamefont {Sanchez-Garcia}}, \ and\
  \bibinfo {author} {\bibfnamefont {H.~F.}\ \bibnamefont {Bettinger}},\
  }\href@noop {} {\bibfield  {journal} {\bibinfo  {journal} {Angew. Chem.}\
  }\textbf {\bibinfo {volume} {57}},\ \bibinfo {pages} {10506} (\bibinfo {year}
  {2018})}\BibitemShut {NoStop}%
\bibitem [{\citenamefont {Yang}\ \emph {et~al.}(2016)\citenamefont {Yang},
  \citenamefont {Davidson},\ and\ \citenamefont {Yang}}]{acene-TDDFT}%
  \BibitemOpen
  \bibfield  {author} {\bibinfo {author} {\bibfnamefont {Y.}~\bibnamefont
  {Yang}}, \bibinfo {author} {\bibfnamefont {E.~R.}\ \bibnamefont {Davidson}},
  \ and\ \bibinfo {author} {\bibfnamefont {W.}~\bibnamefont {Yang}},\
  }\href@noop {} {\bibfield  {journal} {\bibinfo  {journal} {Proc. Natl. Acad.
  Sci. U.S.A.}\ }\textbf {\bibinfo {volume} {113}},\ \bibinfo {pages} {E5098}
  (\bibinfo {year} {2016})}\BibitemShut {NoStop}%
\bibitem [{\citenamefont {Siebrand}(1967)}]{Naphexp}%
  \BibitemOpen
  \bibfield  {author} {\bibinfo {author} {\bibfnamefont {W.}~\bibnamefont
  {Siebrand}},\ }\href@noop {} {\bibfield  {journal} {\bibinfo  {journal} {J.
  Chem. Phys.}\ }\textbf {\bibinfo {volume} {47}},\ \bibinfo {pages} {2411}
  (\bibinfo {year} {1967})}\BibitemShut {NoStop}%
\bibitem [{\citenamefont {Schriber}\ \emph {et~al.}(2018)\citenamefont
  {Schriber}, \citenamefont {Hannon}, \citenamefont {Li},\ and\ \citenamefont
  {Evangelista}}]{Naphzpe}%
  \BibitemOpen
  \bibfield  {author} {\bibinfo {author} {\bibfnamefont {J.~B.}\ \bibnamefont
  {Schriber}}, \bibinfo {author} {\bibfnamefont {K.~P.}\ \bibnamefont
  {Hannon}}, \bibinfo {author} {\bibfnamefont {C.}~\bibnamefont {Li}}, \ and\
  \bibinfo {author} {\bibfnamefont {F.~A.}\ \bibnamefont {Evangelista}},\
  }\href@noop {} {\bibfield  {journal} {\bibinfo  {journal} {J. Chem. Theory
  Comput.}\ }\textbf {\bibinfo {volume} {14}},\ \bibinfo {pages} {6295}
  (\bibinfo {year} {2018})}\BibitemShut {NoStop}%
\bibitem [{\citenamefont {Sharma}\ \emph {et~al.}(2019)\citenamefont {Sharma},
  \citenamefont {Bernales}, \citenamefont {Knecht}, \citenamefont {Truhlar},\
  and\ \citenamefont {Gagliardi}}]{LauraDMRG}%
  \BibitemOpen
  \bibfield  {author} {\bibinfo {author} {\bibfnamefont {P.}~\bibnamefont
  {Sharma}}, \bibinfo {author} {\bibfnamefont {V.}~\bibnamefont {Bernales}},
  \bibinfo {author} {\bibfnamefont {S.}~\bibnamefont {Knecht}}, \bibinfo
  {author} {\bibfnamefont {D.~G.}\ \bibnamefont {Truhlar}}, \ and\ \bibinfo
  {author} {\bibfnamefont {L.}~\bibnamefont {Gagliardi}},\ }\href@noop {}
  {\bibfield  {journal} {\bibinfo  {journal} {Chem. Sci.}\ }\textbf {\bibinfo
  {volume} {10}},\ \bibinfo {pages} {1716} (\bibinfo {year}
  {2019})}\BibitemShut {NoStop}%
\bibitem [{\citenamefont {Hajgató}\ \emph {et~al.}(2011)\citenamefont
  {Hajgató}, \citenamefont {Huzak},\ and\ \citenamefont {Deleuze}}]{FPACCSDT}%
  \BibitemOpen
  \bibfield  {author} {\bibinfo {author} {\bibfnamefont {B.}~\bibnamefont
  {Hajgató}}, \bibinfo {author} {\bibfnamefont {M.}~\bibnamefont {Huzak}}, \
  and\ \bibinfo {author} {\bibfnamefont {M.~S.}\ \bibnamefont {Deleuze}},\
  }\href@noop {} {\bibfield  {journal} {\bibinfo  {journal} {J. Phys. Chem. A}\
  }\textbf {\bibinfo {volume} {115}},\ \bibinfo {pages} {9282} (\bibinfo {year}
  {2011})}\BibitemShut {NoStop}%
\bibitem [{\citenamefont {Ibeji}\ and\ \citenamefont {Ghosh}(2015)}]{NaphGeo}%
  \BibitemOpen
  \bibfield  {author} {\bibinfo {author} {\bibfnamefont {C.~U.}\ \bibnamefont
  {Ibeji}}\ and\ \bibinfo {author} {\bibfnamefont {D.}~\bibnamefont {Ghosh}},\
  }\href@noop {} {\bibfield  {journal} {\bibinfo  {journal} {Phys. Chem. Chem.
  Phys.}\ }\textbf {\bibinfo {volume} {17}},\ \bibinfo {pages} {9849} (\bibinfo
  {year} {2015})}\BibitemShut {NoStop}%
\bibitem [{\citenamefont {Boyn}\ \emph {et~al.}(2020)\citenamefont {Boyn},
  \citenamefont {Xie}, \citenamefont {Anderson},\ and\ \citenamefont
  {Mazziotti}}]{CoDimer}%
  \BibitemOpen
  \bibfield  {author} {\bibinfo {author} {\bibfnamefont {J.-N.}\ \bibnamefont
  {Boyn}}, \bibinfo {author} {\bibfnamefont {J.}~\bibnamefont {Xie}}, \bibinfo
  {author} {\bibfnamefont {J.~S.}\ \bibnamefont {Anderson}}, \ and\ \bibinfo
  {author} {\bibfnamefont {D.~A.}\ \bibnamefont {Mazziotti}},\ }\href@noop {}
  {\bibfield  {journal} {\bibinfo  {journal} {J. Phys. Chem. Lett.}\ }\textbf
  {\bibinfo {volume} {11}},\ \bibinfo {pages} {4584} (\bibinfo {year}
  {2020})}\BibitemShut {NoStop}%
\bibitem [{\citenamefont {Sharma}\ \emph {et~al.}(2020)\citenamefont {Sharma},
  \citenamefont {Truhlar},\ and\ \citenamefont {Gagliardi}}]{LauraDimer}%
  \BibitemOpen
  \bibfield  {author} {\bibinfo {author} {\bibfnamefont {P.}~\bibnamefont
  {Sharma}}, \bibinfo {author} {\bibfnamefont {D.~G.}\ \bibnamefont {Truhlar}},
  \ and\ \bibinfo {author} {\bibfnamefont {L.}~\bibnamefont {Gagliardi}},\
  }\href@noop {} {\bibfield  {journal} {\bibinfo  {journal} {J. Am. Chem.
  Soc.}\ }\textbf {\bibinfo {volume} {142}},\ \bibinfo {pages} {16644}
  (\bibinfo {year} {2020})}\BibitemShut {NoStop}%
\end{thebibliography}%

\end{document}